\documentclass[a4paper,fleqn,usenatbib]{mnras}
\usepackage{newtxtext,newtxmath}
\usepackage[T1]{fontenc}
\usepackage{ae,aecompl}
\usepackage{graphicx}	
\usepackage{amsmath,bm}	
\usepackage{amssymb}

\def\MC{meridional circulation}	
\def\vb{{\bf v}}
\def\Fn{{\bf F}_{\nu}}
\def\FL{{\bf F}_L}
\def\pa{\partial}

\title[Solar cycle variation of meridional circulation]{A theoretical model of the variation of the meridional circulation with the solar cycle}
\author[Hazra \& Choudhuri]{
Gopal Hazra,$^{1,2}$\thanks{E-mail: hgopal@iisc.ac.in}
Arnab Rai Choudhuri$^{1}$
\\
$^{1}$Department of Physics, Indian Institute of Science, Bangalore 560012, India\\
$^{2}$Indian Institute of Astrophysics, Koramangala,Bangalore 560034, India\\
}

\pubyear{2017}

\begin{document}
\label{firstpage}
\pagerange{\pageref{firstpage}--\pageref{lastpage}}
\maketitle

\begin{abstract}
Observations of the meridional circulation of the Sun,
which plays a key role in the operation of the solar dynamo, indicate that its speed  
varies with the solar cycle, becoming faster during the solar
minima and slower during the solar maxima. To explain this variation of the
meridional circulation with the solar cycle, we construct a theoretical model
by coupling the equation of the meridional circulation (the $\phi$ component of the vorticity
equation within the solar convection zone) with the equations of the flux transport dynamo model. 
We consider the back reaction due to the Lorentz force of the dynamo-generated magnetic fields
and study the perturbations produced in the meridional circulation due to it.
This enables us to model the variations of the meridional circulation without developing a full theory of 
the meridional circulation itself. We obtain results which reproduce the observational data of solar cycle variations 
of the meridional circulation reasonably well. We get the best results on assuming the turbulent
viscosity acting on the velocity field to be comparable to the magnetic diffusivity
(i.e.\ on assuming the magnetic Prandtl number to be close to unity).
We have to assume an appropriate bottom boundary condition to ensure that the Lorentz force cannot drive a flow
in the subadiabatic layers below the bottom of the tachocline. Our results are sensitive to this bottom 
boundary condition.  We also suggest a hypothesis how the observed inward flow towards the active
regions may be produced.
\end{abstract}


\section{Introduction}

The meridional circulation is one of the most important large-scale coherent flow
patterns within the solar convection zone, the other such important 
flow pattern being the differential rotation. A strong evidence for the
meridional circulation came when low-resolution magnetograms noted that 
the `diffuse' magnetic field (as it appeared in low resolution) on
the solar surface outside active regions formed unipolar bands which shifted
poleward, implying a poleward flow at the solar surface
\citep{Howard81, WSN89b}. A further confirmation
for such a flow came from the observed poleward migration of filaments
formed over magnetic neutral lines \citep{MFS83,MS89}.
Efforts for direct measurement of the meridional circulation
also began in the 1980s \citep{LH82,Ulrich88}.
The maximum velocity of the meridional circulation at mid-latitudes
is of the order of 20 m s$^{-1}$.

After the development of helioseismology, there were attempts
to determine the nature of the meridional circulation below the
solar surface \citep{Giles97,BF98}.
The determination of the meridional circulation in the lower
half of the convection zone is a particularly challenging problem.
Since the meridional circulation is produced by the turbulent
stresses in the convection zone, most of the flux transport dynamo
models in which the meridional circulation plays a crucial role
assume the return flow of the \MC\ to be at the bottom of the
convection zone.  In the last few years, several authors presented
evidence for a shallower return flow
\citep{Hathaway12, Zhao13, Schad13}. 
However, \citet{RA15} argue that helioseismic
data still cannot rule out the possibility of the return flow
being confined near the bottom of the convection zone.

Helioseismic measurements indicate a variation of the
meridional circulation with the solar cycle.
\citep{CD01,Beck02,BA10,Komm15}.  These results are consistent with
the surface measurements of \citet{Hathaway10b},
who find that the \MC\ at the surface becomes weaker during the sunspot
maximum by an amount of order 5 m s$^{-1}$. One plausible
explanation for this is the back-reaction of the dynamo-generated
magnetic field on the \MC\ due to the Lorentz force.  The aim
of this paper is to develop such a model of the variation of
the \MC\ and to show that the results of such a theoretical
model are in broad agreement with observational data.

The theory of the \MC\ is somewhat complicated.  The turbulent
viscosity in the solar convection zone is expected to be anisotropic,
since gravity and rotation introduce two preferred directions at
any point within the convection zone.  A classic paper by 
\citet{Kippenhahn63} showed that an anisotropic viscosity gives rise
to \MC\ along with differential rotation.  This work neglected the
`thermal wind' that would arise if the surfaces of constant density
and constant pressure do not coincide. A more complete model
based on a mean field theory of convective turbulence is due to
\citet{Kitchatinov95} and extended further by \citet{Kitchatinov11b}. 
These authors argue that, in the high Taylor number situation
appropriate for the Sun, the \MC\ should arise from the small imbalance 
between the thermal wind and the centrifugal force due to the variation
of the angular velocity with $z$ (the direction parallel to the rotation
axis).  See the reviews by Kitchatinov \citep{Kitchatinov11review,Kitchatinov13}
for further discussion. If the \MC\ really comes from a delicate
imbalance between two large terms, one would theoretically expect
large fluctuations in the \MC.  Numerical simulations of convection in
a shell presenting the solar convection zone produce \MC\ which is
usually highly variable in space and time \citep{Karak14,FM15,Passos17}.
It is still not understood why the observed \MC\ is much more
coherent and stable than what such theoretical considerations
would suggest. There are also efforts to understand whether mean
field models can produce a meridional circulation more complicated
than a single-cell pattern \citep{BY17}.

The \MC\ plays a very critical role in the flux transport dynamo
model, which has emerged as a popular theoretical model of the solar
cycle. This model started being developed from the 1990s \citep{WSN91,
CSD95,Durney95,DC99, Nandy02, CNC04} and its current
status has been reviewed by several authors \citep{Chou11,Chou14,
Charbonneau14,Karakreview14}.  The poloidal field in this
model is generated near the solar surface from tilted bipolar sunspot
pairs by the Babcock--Leighton
mechanism and then advected by the \MC\ to higher latitudes
\citep{DC94, DC95,CD99}.
This poleward advection of the poloidal field is an important feature
of the flux transport dynamo model and helps in matching various
aspects of the observational data.  The poloidal field eventually
has to be brought to the bottom of the convection zone, where the
differential rotation of the Sun acts on it to produce the toroidal
field.  If the turbulent diffusivity of the convection zone is assumed
low, then the \MC\ is responsible for transporting the poloidal field
to the bottom of the convection zone.  On the other hand, if the
diffusivity is high, then diffusivity can play the dominant role in
the transportation of the poloidal field across the convection zone
\citep{Jiang07,Yeates08}.
We shall discuss this point in more detail later in the paper. Most
of the authors working on the flux transport dynamo assumed a one-cell
\MC\ with an equatorward flow at the bottom of the convection zone.
This equatorward \MC\ plays a crucial role in the equatorward transport
of the toroidal field generated at the bottom of the convection zone,
leading to solar-like butterfly diagrams.  In the absence of the \MC,
one finds a poleward dynamo wave \citep{CSD95}. When several
authors claimed to find evidence for a shallower \MC, there was a 
concern about the validity of the flux transport dynamo model. \citet{HKC14}
showed that, even if the \MC\ has a much more complicated multi-cell
structure, still we can retain most of the attractive features of the
flux transport dynamo as long as there is a layer of equatorward 
flow at the bottom of the convection zone.  In the present work, we
assume a one-cell \MC.
 
Most of the papers on the flux transport dynamo are of kinematic
nature, in which the only back-reaction of the magnetic field which
is often included is the so-called ``$\alpha$-quenching'', i.e. the
quenching of the generation mechanism of the poloidal field. Usually
kinematic models do not include the back-reaction of the magnetic field
on the large-scale flows. We certainly expect the Lorentz force of the
dynamo-generated magnetic field to react back on the large-scale flows.
From observations, we are aware of two kinds of back-reactions on the
large-scale flows: solar cycle variations of the differential rotation
known as torsional oscillations and solar cycle variations of the \MC\ \citep{Chou11a}.
There have been several calculations showing how torsional oscillations
are produced by the back-reaction of the dynamo-generated magnetic
field \citep{Durney00,Covas00,Bushby06,CCC09}. \citet{Rempel06} presented 
a full mean-field model of both the dynamo and
the large-scale flows. This model produced both torsional oscillations
and the solar cycle variation of the \MC.  However, apart from Figure~4(b)
in the paper showing the time-latitude plot of $v_{\theta}$ near the bottom
of the convection zone, this paper does not study the time variation of the
\MC\ in detail.  To the best of our knowledge, this is the only mean-field
calculation of magnetic back-reaction on the \MC\ before our work. 

The back-reaction on the \MC\ should mainly come from the tension of the
dynamo-generated toroidal magnetic field. Suppose we consider a ring of
toroidal field at the bottom of the
convection zone.  Because of the tension, the ring will try to shrink in
size and this can be achieved most easily by a poleward slip \citep{Van88}. 
It is this tendency of poleward slip that would give
rise to the Lorentz force opposing the equatorward meridional circulation.
We show in our calculations based on a mean field model that this gives rise to a vorticity opposing
the normal vorticity associated with the \MC\ and, as this vorticity diffuses
through the convection zone, there is a reduction in the strength of the \MC\
everywhere including the surface. In order to circumvent the difficulties
in our understanding of the \MC\ itself, we develop a theory of how the Lorentz force
produces a modification of the \MC\ with the solar cycle. The important
difference of our formalism from the formalism of \citet{Rempel06} is the following.
In Rempel's formalism, one requires a proper formulation of turbulent stresses
to calculate the full \MC\ in which one sees a cycle variation due to the 
dynamo-generated Lorentz force.  On the other hand, in our formalism, we have
developed a theory of the modifications of the \MC\ due to the Lorentz force
so that we can study the modifications of the \MC\ without developing
a full theory of the \MC. 

The mathematical formulation of the theory is described in the next Section.
Then Section~\ref{sec:result} presents the results of our simulation.  Our conclusions are
summarized in Section~\ref{sec:conclusion}.

\section{Mathematical Formulation}\label{sec:model}

Our main aim is to study how the Lorentz force due to the dynamo-generated
magnetic field acts on the meridional circulation.  
We need to solve the dynamo equations along with the equation for the \MC.

We write the magnetic field in the form
\begin{equation}
\label{eq:decomp}
{\bf B} = B_\phi(r,\theta,t)\hat{e}_\phi + \nabla\times [A(r,\theta,t) \hat{e}_\phi]
\end{equation}
where $B_\phi$ is the toroidal component of magnetic field and A is the magnetic vector potential 
corresponding to the poloidal component of the field. Then the dynamo equations for
the toroidal and the poloidal components are 
\begin{equation}
\label{eq:Aeq}
\frac{\partial A}{\partial t} + \frac{1}{s}({\bf v}_m.\nabla)(s A)
= \eta_{p} \left( \nabla^2 - \frac{1}{s^2} \right) A + S(r, \theta, t),
\end{equation}

\begin{eqnarray}
\label{eq:Beq}
\frac{\partial B_{\phi}}{\partial t}
+ \frac{1}{r} \left[ \frac{\partial}{\partial r}
(r v_r B_{\phi}) + \frac{\partial}{\partial \theta}(v_{\theta} B_{\phi}) \right]
= \eta_{t} \left( \nabla^2 - \frac{1}{s^2} \right) B_{\phi} \nonumber \\
+ s({\bf B}_p.{\bf \nabla})\Omega + \frac{1}{r}\frac{d\eta_t}{dr}\frac{\partial{(rB_{\phi})}}{\partial{r}},~~~~~~~~~~~~~~~~~~~~
\end{eqnarray}\\
where $v_r$ and $v_{\theta}$ are components of the 
meridional flow $\vb_m$, $\Omega$ is the differential rotation and $s=r \sin\theta$, whereas $S(r, \theta, t)$ is the 
source function which incorporates the Babcock-Leighton mechanism and magnetic buoyancy as explained in \citet{CNC04}
and \citet{CH16}.
We have kept most of the parameters the same as in \citet{CNC04}. In the kinematic approach, both $\vb_m = v_r {\bf e}_r + 
v_{\theta} {\bf e}_{\theta}$ and $\Omega$
are assumed to be given.  

To understand the effect of the Lorentz force on the meridional circulation, we need to consider
the Navier-Stokes equation with the Lorentz force term, which is
\begin{equation}
\frac{\partial {\bf v}}{\partial t} + ({\bf v}.\nabla){\bf v} = -\frac{1}{\rho}\nabla p + {\bf g} + 
\FL + \Fn (\vb)
\label{eq:basic}
\end{equation} 
where ${\bf v}$ is the total plasma velocity, $\bf g$ is the force due to gravity, $\FL$ is the Lorentz
force term and 
$\Fn (\vb)$ is the turbulent viscosity term corresponding to the velocity
field $\vb$. We would have $\Fn (\vb)= \nu \nabla^2 {\bf v}$ if the turbulent
viscosity $\nu$ is assumed constant in space, but $\Fn (\vb)$ can be more complicated for spatially
varying $\nu$. For simplicity we have considered turbulent viscosity as a scalar 
quantity but, in reality, it is a tensor and can be a quite complicated quantity: see \citet{Kitchatinov11b}.
It may be noted that the tensorial nature of turbulent viscosity is quite crucial in the 
theory of the unperturbed \MC\ and it cannot be treated as a scalar in the complete theory.
To study torsional oscillations, we have to consider the $\phi$ component of Eq.~\ref{eq:basic} as done by
\citet{CCC09}.  To consider variations of the \MC, however, we need to focus our attention on
the $r$ and $\theta$ components of Eq.~\ref{eq:basic}. A particularly convenient approach is to take the curl
of Eq.~\ref{eq:basic} (noting that $({\bf v}.\nabla){\bf v} = - \vb \times (\nabla \times \vb) + \frac{1}{2}\nabla (v^2)$)
and consider the $\phi$ component of the resulting equation.  This gives
\begin{eqnarray}
\label{eq:main}
\frac{\partial \omega_{\phi}}{\partial t} - [\nabla \times \{ \vb \times (\nabla \times \vb)\}]_{\phi} = 
\frac{1}{\rho^2}[\nabla \rho \times \nabla p]_{\phi} \\ ~~\nonumber
+ [\nabla \times \FL]_{\phi} + [\nabla \times \Fn(\vb_m)]_{\phi}. 
\end{eqnarray} 
where $\omega = \nabla \times \vb$ is the vorticity, of which the $\phi$ component comes from the 
meridional circulation $\vb_m$ only. 
Writing $\vb = \vb_m + r \sin \theta \Omega\hat{e}_\phi$, a few steps of straightforward algebra give
\begin{equation}
[\nabla \times \{ \vb \times (\nabla \times \vb)\}]_{\phi} = - r \sin\theta\nabla.\left(\vb_m \frac{\omega_{\phi}}{r \sin\theta}\right)
+ r \sin\theta\frac{\partial \Omega^2}{\partial z}
\end{equation}
Substituting this in Eq.~\ref{eq:main}, we have
\begin{eqnarray}
\label{eq:main1}
\frac{\partial \omega_{\phi}}{\partial t} + s \nabla.\left(\vb_m \frac{\omega_{\phi}}{s}\right) =  
s \frac{\partial \Omega^2}{\partial z}+ \frac{1}{\rho^2}(\nabla \rho \times \nabla p)_{\phi} \\ ~~\nonumber
+ [\nabla \times \FL]_{\phi} + [\nabla \times \Fn(\vb_m)]_{\phi}.
\end{eqnarray} 

We now break up the meridional velocity into two parts:
\begin{equation}\label{eq:v_tot}
 \vb_m = \vb_0 + \vb_1,
\end{equation}
where $\vb_0$ is the regular meridional circulation the Sun would have in the absence of
magnetic fields and $\vb_1$ is its modification
due to the Lorentz force of the dynamo-generated magnetic field. The azimuthal vorticity $\omega_{\phi}$
can also be broken into two parts corresponding to these two parts of the \MC:
\begin{equation}
 \omega_{\phi} = \omega_0 + \omega_1.
\end{equation}
It is clear from Eq.~\ref{eq:main1} that the regular \MC\ of the Sun in the absence of magnetic fields, which is assumed
independent of time in a mean field model, should be given by
\begin{eqnarray}
\label{eq:w0}
s \nabla.\left(\vb_0 \frac{\omega_0}{s}\right) =  
s \frac{\partial \Omega^2}{\partial z}+ \frac{1}{\rho^2}(\nabla \rho \times \nabla p)_{\phi} \\ ~~\nonumber
+ [\nabla \times \Fn(\vb_0)]_{\phi}.
\end{eqnarray} 
This is the basic equation which has to be solved to develop a theory of the regular \MC.
The first term in R.H.S is the centrifugal force term, whereas  
the second term, which can  be written as
\begin{equation}
\label{eq:s}
\frac{1}{\rho^2}(\nabla \rho \times \nabla p)_{\phi} = -\frac{g}{c_pr}\frac{\partial S}{\partial \theta},
\end{equation}
is the thermal wind term. This is the term which couples the theory to the thermodynamics
of the Sun.  So we need to solve the energy transport equation of the Sun in order to
develop a theory of the regular \MC, making the problem particularly difficult.
In the pioneering study of \citet{Kippenhahn63}, the thermal wind term was neglected. However,
we now realize that for the Sun, in which the Taylor number is large, this term has
to be important and of the order of the centrifugal term \citep{Kitchatinov95}. It may
be mentioned that, in the model of the \MC\ due to \citet{Kitchatinov95} and
\citet{Kitchatinov11b}, the advection term on the L.H.S.\ of Eq.~\ref{eq:w0} was neglected. On inclusion
of this term, the results were found not to change much (L.\ Kitchatinov, private
communication).

We now subtract Equation~\ref{eq:w0} from \ref{eq:main1}, which gives
\begin{eqnarray}
\label{eq:main2}
\frac{\partial \omega_1}{\partial t} + s \nabla. \left(\vb_0 \frac{\omega_1}{s}\right) + 
s \nabla. \left( \vb_1 \frac{\omega_0}{s}\right) \\ ~\nonumber
= [\nabla \times \FL]_{\phi} + [\nabla \times \Fn(\vb_1)]_{\phi}
\end{eqnarray}
on neglecting the quadratic term in perturbed quantities $\vb_1 \omega_1$ and assuming that
$\Fn(\vb)$ is linear in $\vb$, which is the case if we use the standard expressions of the
viscous stress tensor. We have also assumed that the dynamo-generated magnetic fields do not
affect the thermodynamics significantly, making the thermal wind term to drop out of Eq.~\ref{eq:main2}.
This is what makes the theory of the modification of the \MC\ decoupled from the
thermodynamics of the Sun and simpler to handle than the theory of the unperturbed \MC.
We have also not included the centrifugal force term, on the assumption that the temporal variations
in $\Omega$ with the solar cycle do not significantly affect our model of the \MC\ perturbations. The existence
of torsional oscillations indicates that this may not be a fully justifiable assumption.
In a future work, we plan to include torsional oscillations in the theory along
with the \MC\ perturbations.  

\begin{figure*}
\includegraphics[width=0.92\textwidth]{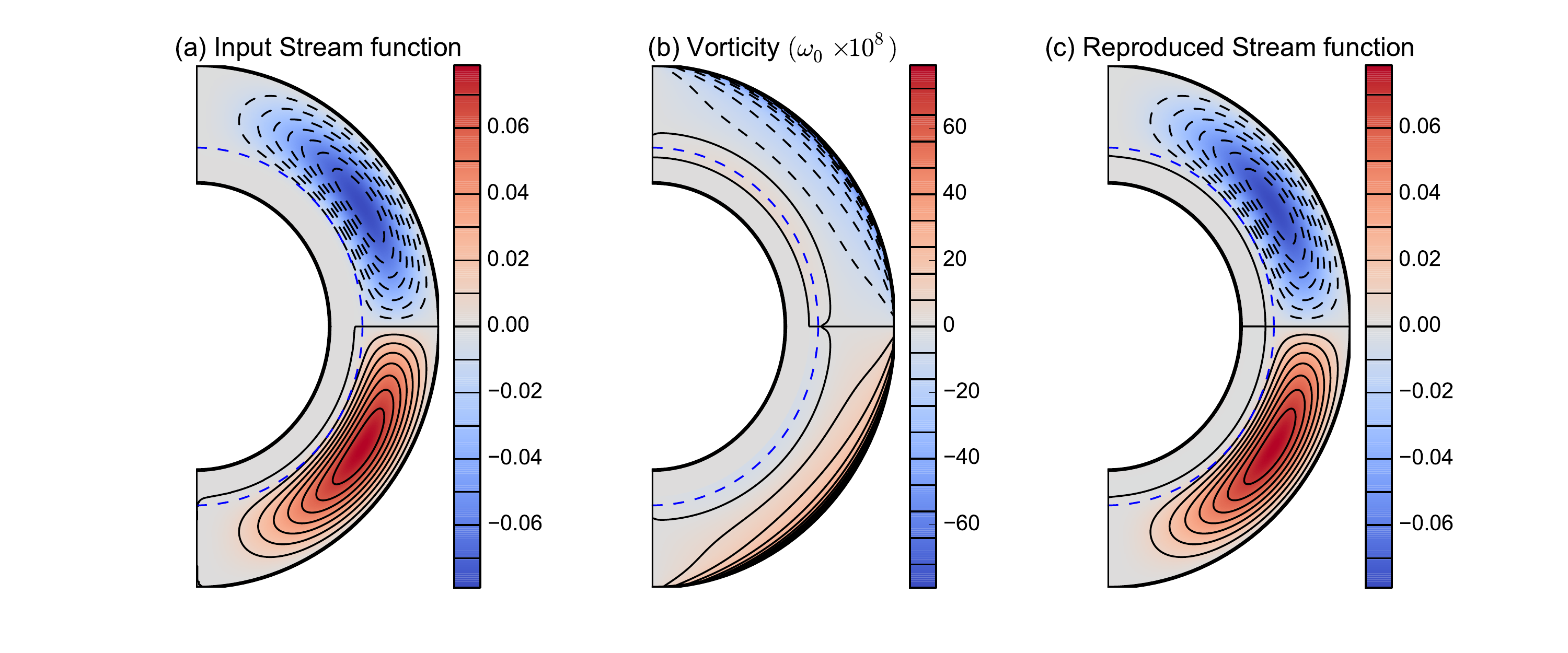}
\caption{(a) The input stream function for the unperturbed meridional circulation which is used in all of our simulations.
The streamlines are indicated by contours---solid contours indicating clockwise flow and dashed contours
indicating anti-clockwise flow.
(b) The unperturbed vorticity calculated from the stream function shown in (a). Solid and dashed
contours indicate positive and negative vorticity. 
(c) The stream function by inverting the vorticity shown in (b) with the help of Eq.~\ref{eq:sv}. This
has to be compared with (a). Blue dashed line represents the tachocline in all plots}
\label{fig:stream_vort}
\end{figure*}

In order to study how the \MC\ evolves with the solar cycle due to the Lorentz force
of the dynamo-generated magnetic fields, we need to solve Equation~\ref{eq:main2} along with
Equations~\ref{eq:Aeq} and \ref{eq:Beq}. A solution of Eq.~\ref{eq:main2} first yields the perturbed vorticity $\omega_1$
at different steps. We can then compute the perturbed velocity $\vb_1$ in the following way.
Since $\nabla\cdot\rho{\bf v}_1 = 0$, we can write $\vb_1$ in terms of a stream
function:
\begin{equation}
 {\bf v}_1 = \frac{1}{\rho} \nabla \times [\psi (r, \theta,t) \hat{e}_\phi].
\end{equation}
Putting this in ${\bf \omega}_1 = \nabla \times {\bf v}_1$, we get
\begin{equation}
\label{eq:sv}
\frac{1}{\rho}[\nabla \rho \times \{\nabla\times ( \psi \hat{e}_\phi)\}]_{\phi} + 
\left(\nabla^2 - \frac{1}{s^2} \right) \psi = -\rho \omega_1.
\end{equation}
This is similar to Poisson's equation and we solve it for a given vorticity to get the stream function. After obtaining
the stream function, it is very trivial to get the velocity fields. There are various methods to solve this
vorticity-stream function equation (Eq.~\ref{eq:sv}), but we choose to solve it using Alternating Direction Implicit (ADI) method. 
The advantage of this method is that it will take less iterations than the other relaxation methods (e.g, Jacobi iteration method) 
to achieve the same accuracy level. To check whether this method is giving us the correct results, 
we take the vorticity of the unperturbed \MC\ that we use in our dynamo calculations. We invert this
vorticity to obtain the velocity field by solving Eq.~\ref{eq:sv} and then cross check if we have succeeded in reproducing
the original input \MC.
The stream function corresponding to the input velocity field is shown in Figure~\ref{fig:stream_vort}(a). Here black solid 
contours surrounding regions of positive stream function (indicated by red)
imply clockwise flow and black dashed contours 
surrounding regions of negative stream function (indicated by blue) imply anti-clockwise flow.
We calculate the vorticity (Fig~\ref{fig:stream_vort}(b)) from this given velocity field and solve Eq.~\ref{eq:sv} 
to get back the stream function. The stream function obtained in this process is shown in Fig~\ref{fig:stream_vort}(c). As we can see, it is almost
indistinguishable from the original given stream function.
While solving Eq. ~\ref{eq:sv}, we have set the tolerance level to $10^{-5}$, which is found
to be sufficient enough to get accurate results. As meridional circulation cannot penetrate much below the tachocline, we have
used $v_r= 0$ at $r=0.65R_\odot$ as our bottom boundary condition.

It may be noted that a realistic specification of density $\rho$ is more important for studying
the \MC\ than for the dynamo problem.  It is the density stratification which determines how strong
the poleward flow at the surface will be compared to the equatorward flow at the bottom of the
convection zone. 
We use a polytropic, hydrostatic, adiabatic stratification for density that matches the standard
solar model quite well \citep{Jones11}:
\begin{equation}\label{eq:rho}
\overline{\rho} = \rho_i ~ \left(\frac{\zeta(r)}{\zeta(r_1)}\right)^n
\end{equation}
where
\begin{equation}
\zeta(r) = c_0 + c_1 \frac{r_2-r_1}{r}
\end{equation}
\begin{equation}
c_0 = \frac{2 \zeta_0 - \beta - 1}{1-\beta}
\end{equation}
\begin{equation}
c_1 = \frac{(1+\beta)(1-\zeta_0)}{(1-\beta)^2}
\end{equation}
and
\begin{equation}
\zeta_0 = \frac{\beta+1}{\beta \exp(N_\rho/n) + 1}  ~~~.
\end{equation}
Here $\rho_i = \rho(r_1) = 0.1788$ g cm$^{-3}$, $\beta = r_1/r_2 = 0.55$ is the aspect ratio,
$n = 1.5$ is the polytropic index, and $N_\rho = 5$ is the number of density scale heights
across the computational domain, which extends from $r_1 = 0.55R_\odot$ to $r_2 = R_\odot$. 

To solve Eq.~\ref{eq:main2}, we need explicit expressions of $[\nabla \times \FL]_{\phi}$ and $[\nabla \times \Fn(\vb_1)]_{\phi}$.
If the magnetic field is given by Eq.~\ref{eq:decomp}, then we have
\begin{eqnarray}\label{eq:curl_fl}
[\nabla \times \FL]_{\phi} = 
\left[\nabla \times \left( \frac{(\nabla \times {\bf B})\times
{\bf B}}{4\pi\rho}\right)\right]_\phi ~~~~~~~~~~~~~~~~~~~~~~~~\\~\nonumber
=\left[\frac{1}{4\pi\rho}\nabla \times \{(\nabla \times {\bf B})\times {\bf B}\}\right]_\phi
- \left[\{(\nabla \times {\bf B})\times {\bf B}\} \times \nabla \left(\frac{1}{4\pi\rho}
\right)\right]_\phi
\end{eqnarray}
It is straightforward to show that
\begin{eqnarray}\label{eq:tot_lf}
\nabla \times \left[\{(\nabla \times {\bf B})\times {\bf B}\}\right]_\phi ~~~~~~~~~~~~~~~~~~~~~~~~~~~~~~~~~~~~~\\ \nonumber
=\left[\frac{1}{r}\frac{\partial}{\partial r}\left(\frac{j_\phi}{s}\right)\frac{\partial (sA)}{\partial \theta}
-\frac{1}{r}\frac{\partial}{\partial \theta}\left(\frac{j_\phi}{s}\right)\frac{\partial (sA)}{\partial r}\right] \\ \nonumber 
+\left[\frac{1}{r}\frac{\partial}{\partial \theta}\left(\frac{B_\phi}{s}\right)\frac{\partial(sB_\phi)}{\partial r}
-\frac{1}{r}\frac{\partial}{\partial r}\left(\frac{B_\phi}{s}\right)\frac{\partial (sB_\phi)}{\partial \theta}\right]
\end{eqnarray}
where $j_\phi = (\nabla \times {\bf B})_\phi$ is $\phi$ component of current density, which is
associated only with the poloidal field.  It is clear from Eq.~\ref{eq:tot_lf} that the Lorentz forces due to
the poloidal field and due to the toroidal field clearly separate out.
In typical flux transport dynamo simulations, the toroidal magnetic field turns out to be about
two orders of magnitude stronger than the poloidal magnetic field.  Since the Lorentz force goes as
the square of the magnetic field, the Lorentz force due to the toroidal field should be about
four orders of magnitude stronger than that due to the poloidal field.
So we include the Lorentz force due to the toroidal field alone in our calculations. We also
have to keep in mind that the magnetic field in the convection zone is highly intermittent with
a filling factor, say $f$.  The Lorentz force is significant only inside the flux tubes and
a mean field value of this has to be included in the equations of our mean field theory. This
point was discussed by \citet{CCC09}, who showed that, if we use mean field values of the magnetic
field in our equations, then the mean field value of the quadratic Lorentz force will involve
a division by $f$. Keeping this in mind, the expression of $[\nabla \times \FL]_{\phi}$ to be
used when solving Equation~\ref{eq:main2} is obtained from Equations~\ref{eq:curl_fl} and \ref{eq:tot_lf}:
\begin{eqnarray}\label{eq:tor_lf}
[\nabla \times \FL]_{\phi} = \frac{1}{4\pi\rho f}\left[\frac{1}{r^2}\frac{\partial}{\partial \theta}
(B_{\phi}^2)
- \frac{\cot\theta}{r}\frac{\partial}{\partial r}(B_{\phi}^2) \right] \\~\nonumber
+ \frac{1}{4\pi\rho^2 f}\frac{d \rho}{d r}\frac{B_\phi}{r \sin\theta}\frac{\partial (\sin\theta B_\phi)}{\partial \theta}
\end{eqnarray} 
The expression for $[\nabla \times \Fn (\vb_1)]_{\phi}$ happens to be rather complicated (even
when the turbulent diffusivity $\nu$ is assumed to be a scalar) and will be discussed
in Appendix~A. We carry on our calculations assuming a simpler form of the viscosity term: 
\begin{equation}
\label{eq:viscous1}
[\nabla \times F_\nu({\bf v_1})]_\phi = \nu\left(\nabla^2-\frac{1}{r^2 \sin^2\theta}\right) \omega_1.
\end{equation}
As we shall show in Appendix~\ref{sec:appendix}, this simple expression of  $[\nabla \times \Fn (\vb_1)]_{\phi}$,
which is easy to implement in a numerical code, captures the effect of viscosity reasonably well.

To study the variations of the \MC\ with the solar cycle, we need to solve Equations~\ref{eq:Aeq}, \ref{eq:Beq} and \ref{eq:main2}
simultaneously, with $[\nabla \times \FL]_{\phi}$ and $[\nabla \times \Fn (\vb_1)]_{\phi}$ in 
Eq.~\ref{eq:main2} given by Eq.~\ref{eq:tor_lf} and Eq.~\ref{eq:viscous1} respectively.  It is 
the term  $[\nabla \times \FL]_{\phi}$ in Eq.~\ref{eq:main2} 
which is the source of the perturbed vorticity $\omega_1$ and gives rise the variations of the \MC\
with the solar cycle. The toroidal field
field $B_{\phi}$ obtained at each time step from Eq.~\ref{eq:Beq} is used for calculating  
$[\nabla \times \FL]_{\phi}$ as given by Eq.~\ref{eq:tor_lf}.

The numerical procedure for solving Equations~\ref{eq:Aeq} and \ref{eq:Beq} has been discussed in detail by \citet{CNC04}.  We solve
Eq.~\ref{eq:main2} also in a similar way.  Note that this equation 
consists of two advection terms in the LHS and one diffusion term along with the source term in
the RHS. It is also basically an advection-diffusion equation similar to Eq.~\ref{eq:Aeq} and \ref{eq:Beq}. 
We solve Equation~\ref{eq:main2} to obtain the axisymmetric perturbed vorticity $\omega_1$ 
in the $r-\theta$ plane of the Sun with $256 \times 256$ grid cells in latitudinal and radial directions.
As in the case of Equations~\ref{eq:Aeq} and \ref{eq:Beq}, we solve Eq.~\ref{eq:main2} by using Alternating Direction Implicit (ADI) method of differencing,
treating the diffusion term through the Crank--Nicholson scheme and the advection term
through the Lax--Wendroff scheme (for more detail please see the guide
of {\it Surya} code which is publicly available upon request. Please send e-mail to {\it arnab@iisc.ac.in}).
We have used $\omega_1=0$ as the boundary condition at the poles $\theta =0$ and $\pi$. 
The radial boundary conditions are also $\omega_1 =0$ on 
the surface $(r = R_\odot)$ and below the tachocline $(r = 0.70R_\odot)$.

We now make one important point.  By solving Eq.~\ref{eq:main2}, we obtain the perturbed vorticity $\omega_1$ at each time 
step.  If we need the perturbed velocity $\vb_1$ also at a time step, then we further need to solve Eq.~\ref{eq:sv}
over our $256 \times 256$ grid points.  If this
is done at every time step, then the calculation becomes computationally very expensive.
It is much easier to run the code if we do not need $\vb_1$ at each time step. In a completely
self-consistent theory, the meridional flow appearing in the dynamo Equations~\ref{eq:Aeq} and \ref{eq:Beq}
should be given by Eq.~\ref{eq:v_tot}, with $\vb_1$ inserted at each time step.  As we pointed out, this
is computationally very expensive.  If $\vb_1$ is assumed small compared to $\vb_0$, then
the dynamo Equations \ref{eq:Aeq} and \ref{eq:Beq} can be solved much more easily by using $\vb_m = \vb_0$.
Additionally, $\vb_1$ appears in the third term on the LHS of Eq.~\ref{eq:main2}. 
We discuss this term in Appendix B, where we argue that our results should not change
qualitatively on not including this term.  If we neglect this term along with using $\vb_m = \vb_0$
in the dynamo equations \ref{eq:Aeq} and \ref{eq:Beq}, then we do not require $\vb_1$ at each time step. The
calculations in the present paper are done by following this approach, which is computationally
much less intensive.  This has enabled us to explore the parameter space of the problem more
easily and understand the basic physics issues.  We are now involved in the much more
computationally intensive calculations in which $\vb_1$ is calculated at regular time intervals
and the dynamo equations \ref{eq:Aeq} and \ref{eq:Beq} are solved by including $\vb_1$ in the meridional flow.
The results of these calculations will be presented in a future paper.

Finally, before presenting our results, we want to point out a puzzle
which we are so far unable to resolve.  Our calculations suggest
that the variations of the meridional circulation should be about
two orders of magnitude larger than what they actually are! Since a larger
$f$ would reduce the Lorentz force given by Eq.~\ref{eq:tor_lf}, we can make the
variations of the meridional circulation to have the right magnitude
only if we take the filling factor $f$ larger than unity, which is
clearly unphysical.  We show that this problem becomes apparent even
in a crude order-of-magnitude estimate. We expect $\omega_1$ to be of order
$V/L$, where $V$ is the perturbed velocity amplitude and $L$ is the
length scale.  Then $\partial \omega_1/\partial t$ has to be of order
$V/LT$.  Taking $T$ to be the solar cycle period and $L$ to be the depth
of the convection zone, a value $V \approx 5$ m s$^{-1}$ gives
\begin{equation}\label{eq:dw1dt}
\left| \frac{\partial \omega_1}{\partial t} \right| \approx 10^{-16}
\rm{s}^{-2}.
\end{equation}
The curl of the Lorentz force is the source of this term and should
be of the same amplitude.  Now the curl of the Lorentz force is of
order $|B_\phi^2 /4 \pi f \rho L^2|$.  In a mean field dynamo model, the
magnetic field should be scaled such that the mean polar field comes
out to have amplitude of order 10 G.  Then the mean toroidal field at
the bottom of the toroidal field turns out to be about $10^3$ G. On
using such a value, the curl of the Lorentz force is of order $10^{-15}/f$
s$^{-2}$. This would become equal to $10^{-16}$
s$^{-2}$ given in Eq. \ref{eq:dw1dt} only if $f$ is of order 10---a completely
unphysical result.  If we take $f$ less than 1 as expected, then the 
theoretical value of the perturbed \MC\ would be much larger than
what is observed. One may wonder if the same problem would be there
in the theory of torsional oscillations.  It turns out that the torsional
oscillations have the same amplitude 5 m s$^{-1}$ as variations in the
meridional circulation.  However, torsional oscillations are driven by
the component $\approx |B_r B_\phi/L|$ of the magnetic stress rather than
the component $\approx |B_\phi^2/L|$ relevant for meridional circulation
variations. Since $|B_r|$ turns out to be about 100 times smaller $|B_\phi|$
in mean field dynamo calculations, things come out quite reasonably in
the theory of torsional oscillations \citep{CCC09}. Here is then the puzzle.
Since the variations in the meridional circulation are driven by the
magnetic stress $\approx |B_\phi^2/L|$ which is about 100 times larger
than the magnetic stress $\approx |B_r B_\phi/L|$ driving torsional
oscillations, one would naively expect the meridional circulation variations
to have an amplitude about 100 times larger than the amplitude of torsional
oscillations.  But why are their observational values found comparable?
In the present paper, we do not attempt to provide any solution to this
puzzle and merely present this as an issue that requires further study.
We point out that, even in non-magnetic situations, some terms in
the equations tend to produce a much faster meridional circulation
than what is observed \citep{Durney96,Dikpati14}. Presumably, such a
fast meridional circulation would upset the thermal balance and, as a
back reaction, a thermal wind force would arise to ensure that such a large
meridional circulation does not take place. It needs to be
investigated whether similar considerations would hold in the case
of magnetic forcing of the meridional circulation as well.  


\begin{figure}
\centering{
\includegraphics[width=0.4\textwidth]{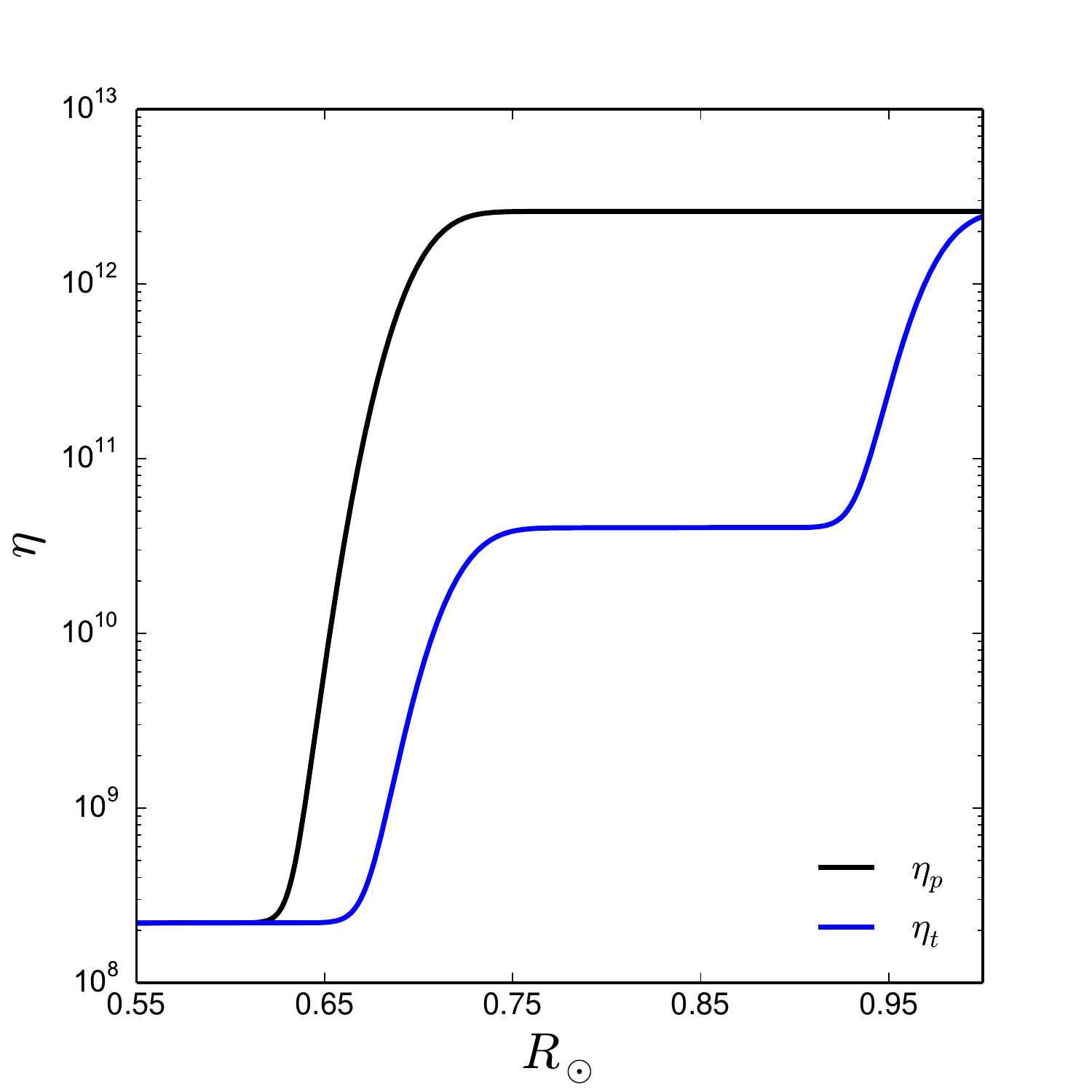}}
\caption{Magnetic diffusivities (cm$^2$ s$^{-1}$) used in our flux transport model. The black solid line shows
the turbulent diffusivity for the poloidal magnetic field, and the blue solid line shows the same for
the toroidal magnetic field.}   
\label{fig:eta}
\end{figure}

\begin{figure}
\centering{
\includegraphics[width=0.45\textwidth]{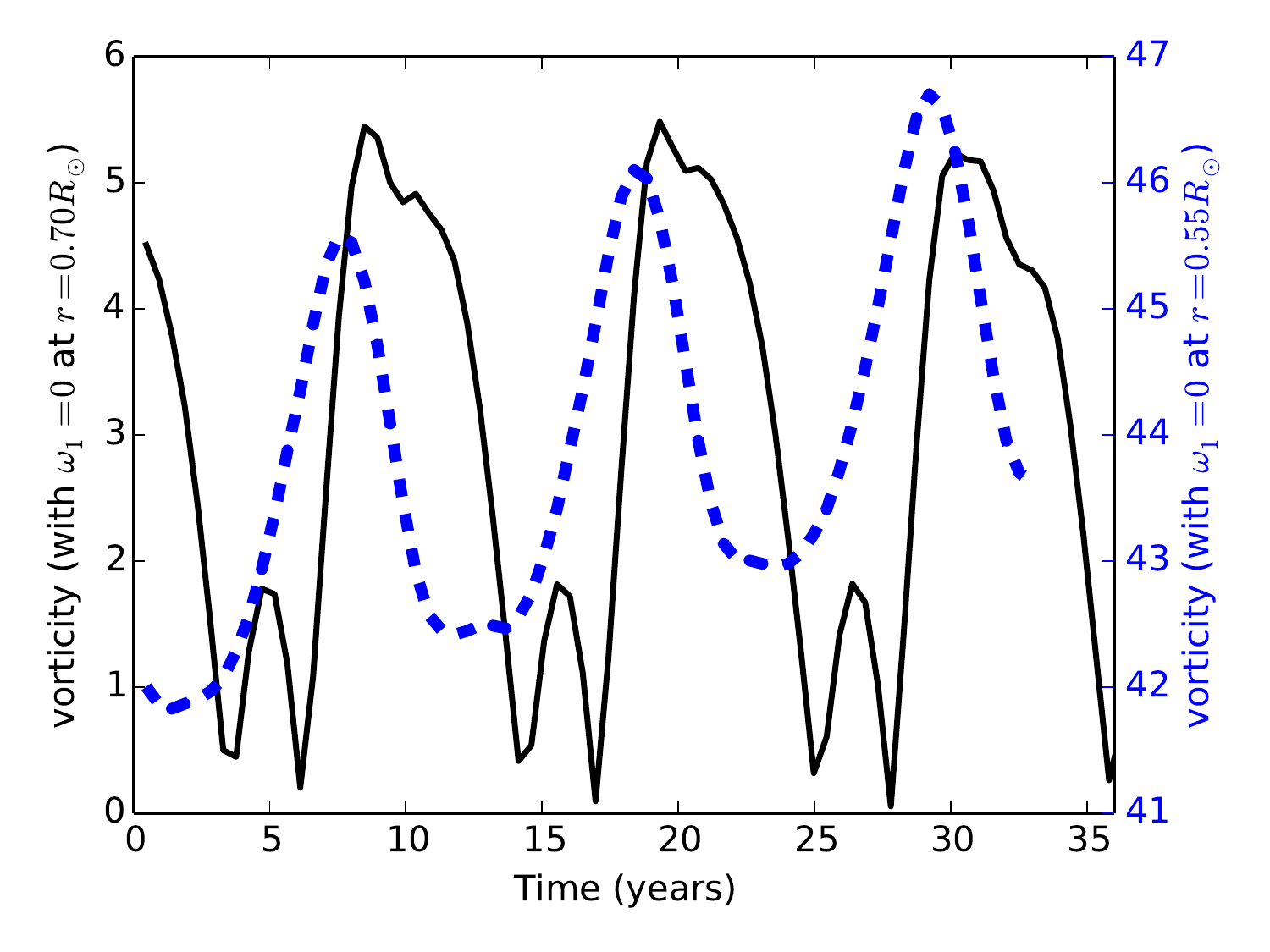}}
\caption{The perturbed vorticity at a point within the convection zone is plotted against time for the two cases
with different boundary conditions. The black solid line shows the absolute value of the perturbed vorticity $\omega_1$ at 
15$^{\circ}$ latitude and at depth $0.75R_\odot$, 
with the  bottom boundary condition $\omega_1 =0$ at $r = 0.7R_\odot$. The blue dashed line shows the absolute value 
of perturbed vorticity $\omega_1$ at $15^{\circ}$ latitude 
and at depth $0.71R_\odot$, with the bottom 
boundary condition $\omega_1 =0$ at $r = 0.55R_\odot$. The values of the vorticity which is shown in blue dashed 
line are given on the right $y$-axis with blue colour.}   
\label{fig:w1_time}
\end{figure}

\begin{figure*}
\centering{
\includegraphics[width=0.84\textwidth]{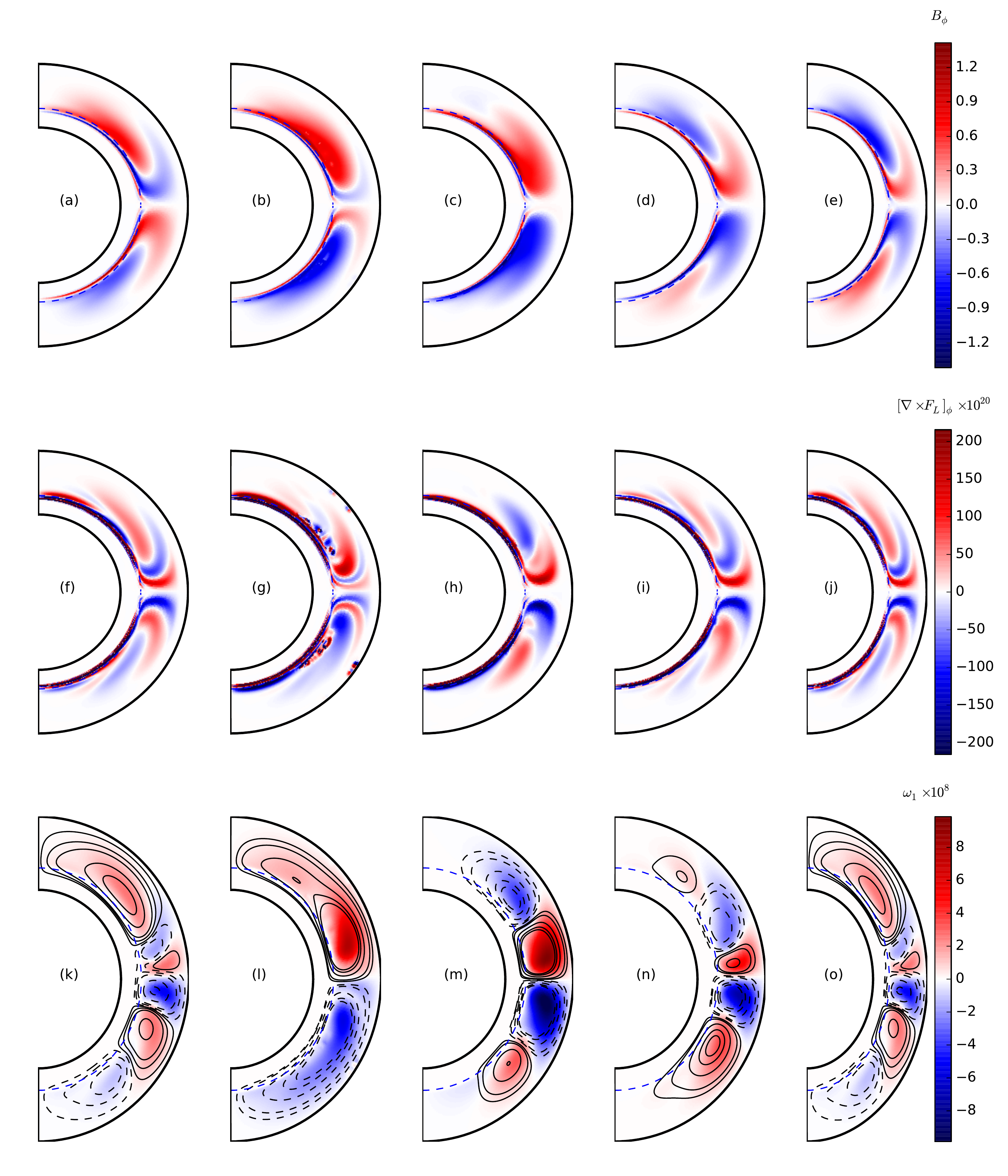}}
\caption{Snapshots in the $r-\theta$ plane of different quantities spanning an entire solar cycle: the toroidal field
$B_{\phi}$, the $\phi$ component of the curl of the Lorentz force $[\nabla \times \FL]_{\phi}$
and the perturbed vorticity $\omega_1$ with streamlines are shown in the first row (a)-(e), the second row (f)-(j), and 
the third row (k)-(o) respectively. The five columns represent time instants during
the midst of the rising phase, the solar maximum, the midst of the decay phase and the solar minimum,
followed by the midst of the rising phase again. 
In the third row (k)-(o), the solid black contours represent clockwise flows and the dashed black contours 
represent anti-clockwise flows. The unit of toroidal fields in the first row is given in terms of
$B_c$. The Lorentz forces in the second row also are calculated from toroidal fields in the 
unit of $B_c$. The vorticities are given (third row) in s$^{-1}$. Here $B_c$ is the critical strength 
of magnetic fields above which the fields are magnetically buoyant and create sunspots \citep{CNC04}.
All the plots are for the magnetic Prandtl number $P_m = 1$ with the bottom boundary condition $\omega_1 =0$ at
$r = 0.7 R_{\odot}$.}   
\label{fig:snap_1d12}
\end{figure*}

\begin{figure*}
\includegraphics[width=1.0\textwidth]{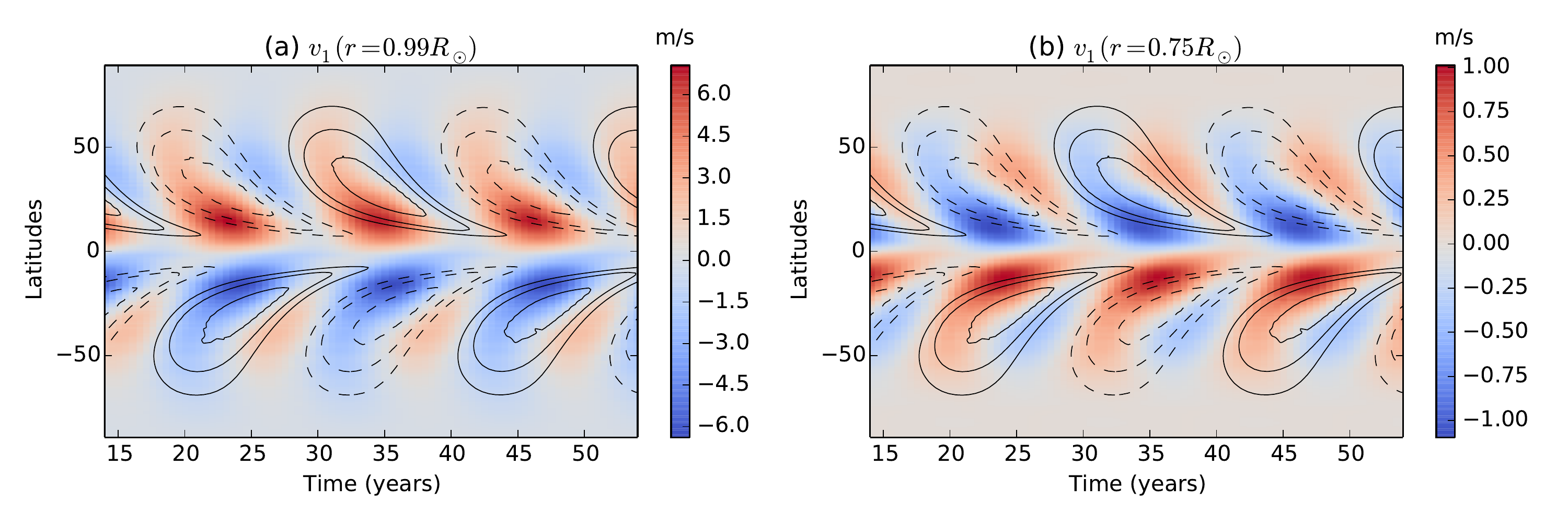}
\caption{The time--latitude plot of perturbed velocities $v_\theta$ (a) near the surface $(r = 0.99R_\odot)$ and 
(b) at the depth $r = 0.75R_\odot$. 
In the northern hemisphere, positive red colour shows perturbed flows towards the equator and the negative blue colour 
shows flows towards pole. It is opposite in the
southern hemisphere. Toroidal fields at the bottom of the convection zone are overplotted in both of the plots 
by line contours. Black solid contours indicate positive polarity and black dashed contours are for negative polarity.
All the plots are for the magnetic Prandtl number $P_m = 1$ with the bottom boundary condition $\omega_1 =0$ at
$r = 0.7 R_{\odot}$.}
\label{fig:bfly_1d12}
\end{figure*}

\begin{figure}
\centering{
\includegraphics[width=0.5\textwidth]{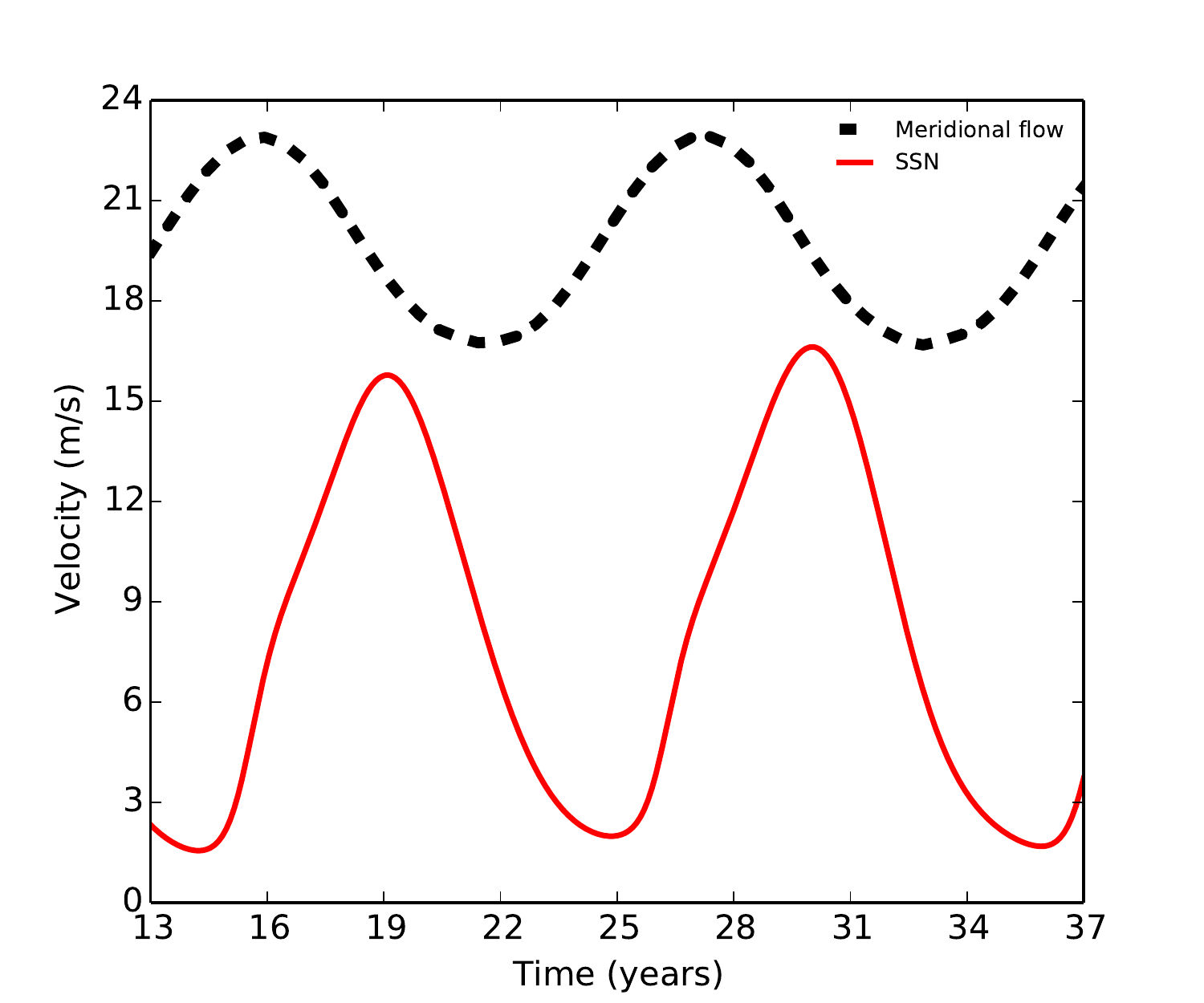}}
\caption{The variation of the total meridional circulation with time 
is plotted with solar cycle. The yearly averaged solar cycle is plotted in the red solid line, whereas the black dashed line shows
the total meridional circulation just below the surface at $25^{\circ}$ latitude.}   
\label{fig:tot_mc}
\end{figure}

\section{Results}\label{sec:result}

We now present results obtained by solving the dynamo equations ~\ref{eq:Aeq} and \ref{eq:Beq}
simultaneously with Eq.~\ref{eq:main2} for perturbed vorticity.
The various dynamo parameters---the source function $S(r, \theta, t)$, the differential rotation
$\Omega$, the meridional circulation $\vb_m$ and the turbulent diffusivities $\eta_p$, $\eta_t$---are
specified as in \citet{CNC04}.  As we have pointed out, calculations in this paper are based
on using the unperturbed \MC\ $\vb_0$ in the dynamo equations ~\ref{eq:Aeq} and \ref{eq:Beq}.
We also follow \citet{CNC04} in
assuming different diffusivity for the poloidal and the toroidal components of the magnetic field as shown
in Fig.~\ref{fig:eta}. The justification for the reduced diffusivity of the toroidal component is that
it is much stronger than the poloidal component and the effective turbulent diffusivity experienced by it is expected
to be reduced by the magnetic quenching of turbulence.  The unquenched value of turbulent diffusivity used 
for the poloidal field inside the convection zone is what is expected from simple mixing length arguments
(\citet{Parker79}, p.\ 629). A much smaller value of turbulent diffusivity was assumed in some other flux transport dynamo
models \citep{DC99}. However, a higher value appears much more realistic on the ground that only with
such a higher value it is possible to model many different aspects of observational data: such as the 
dipolar parity \citep{CNC04,Hotta10}, the lack of significant hemispheric asymmetry \citep{CC06,GoelChou09}, the correlation of the cycle strength with the polar field
during the previous minimum \citep{Jiang07} and the Waldmeier effect \citep{Karakchou11}. 
For solving the vorticity Equation~\ref{eq:main2} we need to specify the turbulent viscosity $\nu$.
From simple mixing length arguments, we would expect the turbulent magnetic diffusivity $\eta_p$
and the turbulent viscosity $\nu$ to be comparable, i.e, we would expect the magnetic Prandtl number $P_m=\nu/\eta_p$ to be close
to 1. We first present results obtained with $P_m =1$ in Section~\ref{sec:pm_1}. How our results get modified 
on using a lower value of the magnetic Prandtl number $P_m$ 
will be discussed in the Section~\ref{sec:lowpm}.

We know that the toroidal field $B_{\phi}$ changes its sign from one cycle to the next.  Since
the Lorentz force depends on the square of $B_{\phi}$, as seen in Eq.~\ref{eq:tor_lf}, the Lorentz
force should have the same sign in different cycles.  As a result, one may think that
the perturbed vorticity ${\bf \omega}_1$
driven by the Lorentz force will tend to grow. However, results presented in Sections~\ref{sec:pm_1} and
\ref{sec:lowpm} are obtained with the boundary condition $\omega_1 = 0$ at $r= 0.7 R_{\odot}$, which ensures that
$\omega_1$ cannot grow indefinitely. Asymptotically, we find $\omega_1$ to have oscillations
as the Lorentz force
increases during solar maxima and decreases during solar minima. In Fig.~\ref{fig:w1_time} we have shown
the time variation of perturbed vorticity (at $15^{\circ}$ latitude and $r = 0.75R_\odot$) for this case with the black solid line.
The calculations presented in Sec.~\ref{sec:lowbc} were done by taking the boundary condition $\omega_1 = 0$ at $r= 0.55 R_{\odot}$.
In this situation, we found that $\omega_1$ has oscillations around a mean which keeps growing
with time (see the blue dashed line plot of Fig.~\ref{fig:w1_time} which does not saturate even after
running the code for several cycles).

\begin{figure*}
\centering{
\includegraphics[width=0.84\textwidth]{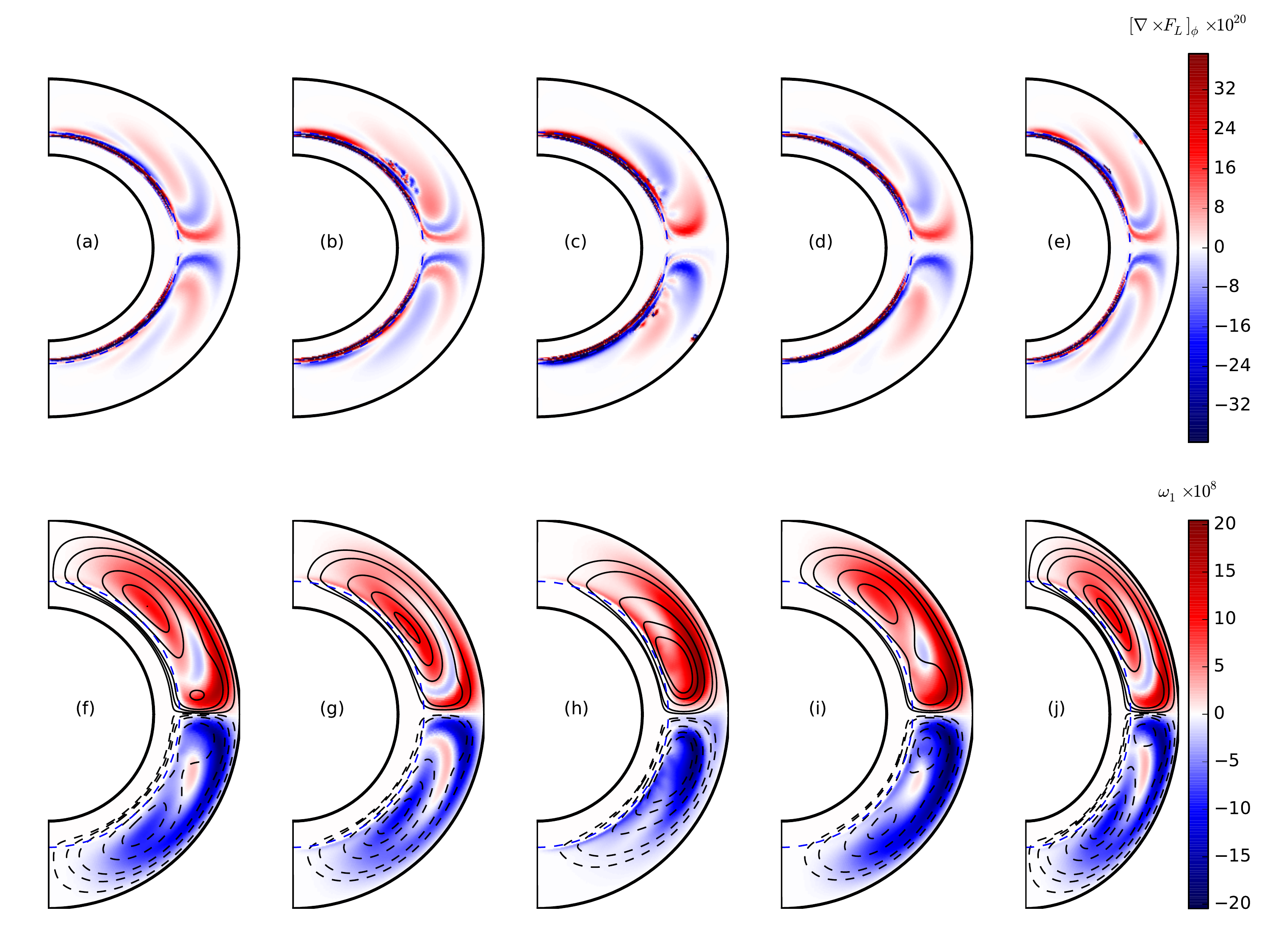}}
\caption{Results obtained with low Prandtl number plotted in the same way as Fig.~\ref{fig:snap_1d12}.
The top row plotting $[\nabla \times \FL]_{\phi}$ is the same as the second row of Fig.~\ref{fig:snap_1d12}.
The bottom row shows the vorticity $\omega_1$ with streamlines.}   
\label{fig:snap_5d10}
\end{figure*}

\begin{figure*}
\centering{
\includegraphics[width=1.0\textwidth]{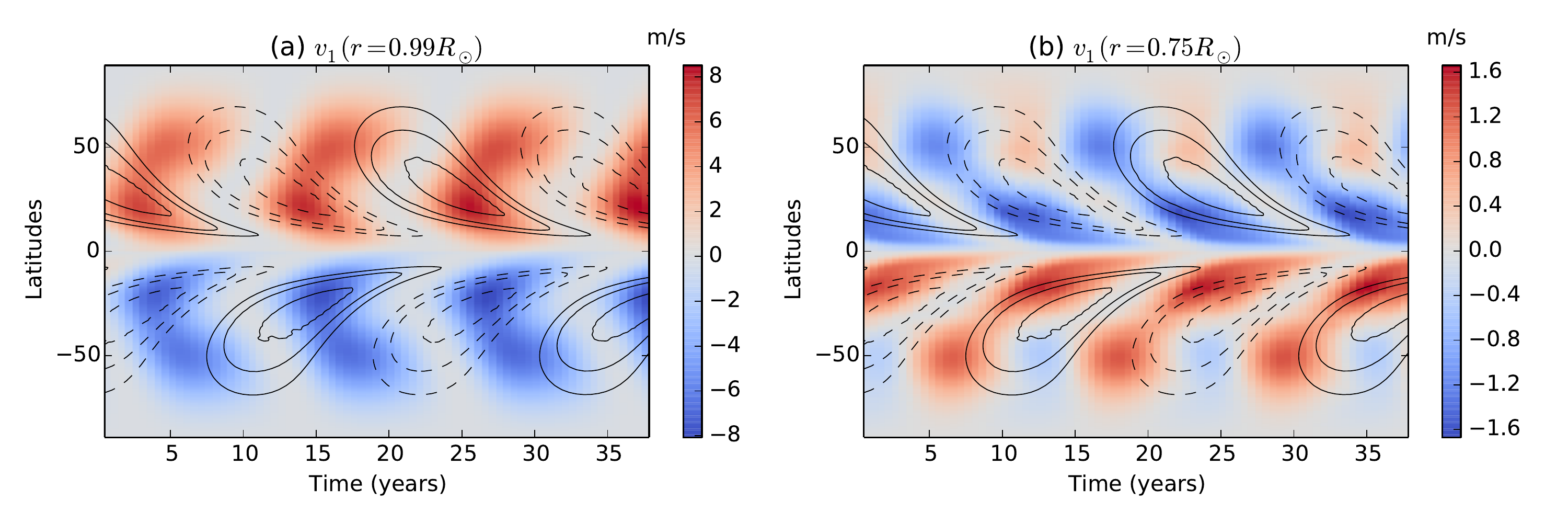}}
\caption{Same as Fig.~\ref{fig:bfly_1d12} but with low Prandtl number $P_m$.}   
\label{fig:bfly_5d10}
\end{figure*}

\subsection{Results with $P_m = 1$}\label{sec:pm_1}
The meridional circulation in a poloidal cut of the Sun's northern hemisphere is anti-clockwise, implying
a negative vorticity $\omega_0$. If the meridional circulation is to be weakened by the Lorentz
forces at the time of the sunspot maximum, then we need to generate a positive perturbed vorticity
$\omega_1$ at that time. Fig.~\ref{fig:snap_1d12} explains the basic physics of how this happens.  The three rows
of this figure plot three quantities in the $r-\theta$ plane: the toroidal field $B_\phi$, the 
$\phi$-component of the curl of the Lorentz force $[\nabla \times {\bf F}_L]_{\phi}$ and the perturbed
vorticity $\omega_1$ along with the associated streamlines. The five vertical columns correspond to five
time intervals during a solar cycle.  The second column corresponds to a time close to the sunspot
maximum, whereas the the fourth column corresponds to the sunspot minimum. 

Our discussion of Figure~\ref{fig:snap_1d12} below will refer to the part in the northern hemisphere.
Signs will have to be opposite for same quantities in the southern hemisphere.
We see in the top row that $B_{\phi}$, produced primarily in the tachocline by the strong
differential rotation there, is concentrated in a layer at the bottom of the solar convection
zone. Since $- \partial (B_{\phi}^2)/ \partial r$ will be positive at the top of this layer
and will be negative at the bottom of this layer, we expect from Eq.~\ref{eq:tor_lf} that $[\nabla \times {\bf F}_L]_{\phi}$
will also respectively have positive and negative values in these regions. This expectation in borne out
as seen in the second row of Fig.~\ref{fig:snap_1d12} plotting $[\nabla \times {\bf F}_L]_{\phi}$. Where the layer
of the toroidal field ends at low latitudes, we have negative $\partial (B_{\phi}^2)/ \partial \theta$
and we see from Eq.~\ref{eq:tor_lf} that this will lead to negative $[\nabla \times {\bf F}_L]_{\phi}$.  This is
also seen in the second row of Fig.~\ref{fig:snap_1d12}. We now note from Eq.~\ref{eq:main2} that $[\nabla \times {\bf F}_L]_{\phi}$
is the main source of the perturbed vorticity $\omega_1$. We expect that the positive $[\nabla \times {\bf F}_L]_{\phi}$
above the tachocline (i.e.\
at the top of the layer of concentrated toroidal field) will give rise to positive $\omega_1$,
implying a clockwise flow opposing the regular \MC. We see in the third row (k)-(o) of Fig.~\ref{fig:snap_1d12} that the
distribution of $\omega_1$ inside the convection zone follows the distribution of 
$[\nabla \times {\bf F}_L]_{\phi}$ (as shown in the second row (f)-(j)) fairly closely. We see that
positive $\omega_1$ is particularly dominant within the convection zone near the time of
the sunspot maximum (second column). This corresponds to streamlines with clockwise
flow, which will oppose the regular \MC\ at the time of the 
sunspot maximum and will lead to a decrease in the overall amplitude of the \MC\ at the solar
surface, in accordance with the observational data. It may be interesting to compare Fig.~\ref{fig:snap_1d12} 
with fig.~3 of \citet{Passos17}, where they present some results on the variation of the meridional circulation
with the solar cycle on the basis of a full numerical simulation.  Our results
obtained from a mean field theory show a spatially smoother perturbed velocity
field (perhaps in better agreement with observational data?) than what is found in the
full simulation.

We wish to make one important point here. A careful look at the second row of Fig.~\ref{fig:snap_1d12} shows
that there is a thin layer of negative $[\nabla \times {\bf F}_L]_{\phi}$ at the bottom of the
tachocline where $- \partial (B_{\phi}^2)/ \partial r$ is negative.  This
layer of negative $[\nabla \times {\bf F}_L]_{\phi}$ will try to create a negative
vorticity driving a counter-clockwise flow below the tachocline.  However, we know that
it is not easy to drive a flow below the tachocline where the temperature gradient is
stable \citep{GM04}. The physics of this is taken into account by using
the bottom boundary condition that $\omega_1 = 0$ at $r = 0.7 R_{\odot}$.  As we shall see
in Section~\ref{sec:lowbc}, results can change dramatically on changing this boundary condition. Our
boundary condition ensures that the negative $[\nabla \times {\bf F}_L]_{\phi}$ at
the bottom of the tachocline cannot produce negative vorticity there (as seen in the 
third row of Fig.~\ref{fig:snap_1d12}) and cannot drive a flow below the tachocline. We thus see that,
while the distribution of $\omega_1$ follows the distribution
of $[\nabla \times {\bf F}_L]_{\phi}$ fairly closely within the convection zone, this
is not the case at the bottom of the tachocline.

The strength of the perturbed \MC\ certainly depends on the strength of $[\nabla \times {\bf F}_L]_{\phi}$,
which is controlled by the filling factor $f$ as seen in Eq.~\ref{eq:tor_lf}. 
We have followed the dynamo model of \citet{CNC04} in which the scale of the magnetic
field is set by the critical value $B_c$ above which the toroidal field becomes buoyant
in the convection zone.  As pointed out by \citet{Jiang07} and followed by \citet{CCC09}, we take
$B_c = 108$ G to ensure that the mean polar field has the right amplitude. With the unit of
the magnetic field fixed in this way, we find that $f = 53$ gives
surface values of the perturbed \MC\ of the order of 5 m s$^{-1}$ comparable to what is seen in the observational
data. Such a value of the filling factor $f$, which should be less than 1, is completely
unphysical.  We anticipated this problem on the basis of the order-of-magnitude estimate
presented in Section~2. 
\citet{CCC09} needed $f = 0.067$ to model torsional
oscillations, whereas \citet{Chou03} estimated that it should be about $f\approx 0.02$.
Since $f$ appears in the denominator of the Lorentz force term in Eq.~\ref{eq:tor_lf},
a large value of $f$ means that we have to reduce the Lorentz force in an artificial manner
to match theory with observations. We are not sure about the significance of this. All we
can say now is that, if we reduce the Lorentz force in this manner, then we find theoretical
results to be in good agreement with observational data.

Fig.~\ref{fig:bfly_1d12} shows time-latitude plots of the perturbed $v_\theta$ just below the surface and in
the lower part of the convection zone. Fig.~\ref{fig:bfly_1d12}(a) can be directly compared with Fig.~11 of
\citet{Komm15}. When the perturbed \MC\ near the surface has an amplitude of about 
5 m s$^{-1}$ near the surface (comparable to what we find in observational data),
its amplitude at the bottom of the convection zone turns out to be about 0.75 m s$^{-1}$.
This is a significant fraction of the amplitude of the \MC\ at the bottom of the convection
zone, which is of order 2 m s$^{-1}$. In other words, the ratio of the amplitude at the 
bottom of the convection zone to the amplitude at the top of the convection zone for the 
perturbed \MC\ ($\approx 0.15$) is somewhat larger than this ratio for the regular \MC
($\approx 0.08$). The reason behind this
becomes clear on comparing the distribution of $\omega_0$, as seen in Fig.~\ref{fig:stream_vort}(b), with the 
distribution of $\omega_1$, as seen in Fig.~\ref{fig:snap_1d12}(l), i.e.\
the second figure in the third row of Fig.~\ref{fig:snap_1d12}. 
This ratio is small when the vorticity is concentrated near the upper part of the 
convection zone and streamlines in the lower part of the convection zone are
fairly spread out.  On the other hand, with $\omega_1$ distributed throughout the body
of the convection zone, this ratio for the perturbed \MC\ is not so small.  
The perturbed \MC\ at the bottom of the convection zone (which is in the poleward
direction) will reduce the value of the equatorward \MC\ there considerably at the
time of the sunspot maximum.  This will presumably affect the behaviour of the dynamo, in
such ways as lengthening its period (since a slower \MC\ at the bottom of the convection
zone during certain phases of the cycle is expected to lengthen the cycle). We are certainly
not fully justified in solving the dynamo equations~\ref{eq:Aeq}--\ref{eq:Beq} by using the regular
\MC\ $\vb_0$.  However, as we have pointed out in Section~\ref{sec:model}, in order to include the perturbed \MC\
in these equations, we need to evaluate the perturbed velocity field at every time step (i.e.\
not only the perturbed vorticity $\omega_1$), which will require considerably more
computational efforts.  We are currently doing these calculations and will present the results
in a forthcoming paper.  The aim of the present exploratory paper is to elucidate the 
basic physics of the problem and not to construct a very complete model. Figure~\ref{fig:tot_mc} shows
the time evolution of the total $v_{\theta}$ at $25^{\circ}$ degree latitude just below the surface
along with the sunspot number.  We see that the \MC\ reaches its minimum a little after the
sunspot maximum. This figure can be compared with Fig.~4 of \citet{Hathaway10b}.

One important aspect of observational data is an inward flow towards the belt of active
regions at the time of the sunspot maximum \citep{CS10,Komm15}.
We now come to the question whether our model can provide any explanation for this. This
inward flow means that the \MC\ is enhanced in the low latitudes below the sunspot belt
and reduced in the higher latitudes. It is thought that the cooling effect of sunspots
may drive this inward flow \citep{Spruit03,Gizon08}.  Since this idea is not
yet fully established through detailed and rigorous calculations, it is worthwhile to look for 
alternative explanations. We would like to draw the reader's attention to the first figure
in the bottom row of Fig.~\ref{fig:snap_1d12}. In the northern hemisphere, we see positive $\omega_1$ (implying
clockwise flow) extending over latitudes higher than where sunspots are usually seen.
However, we see a region of negative $\omega_1$ (implying counter-clockwise flow) at lower
latitudes.  It is clear that the perturbed \MC\ is of the nature of an inward flow at the 
latitudes between the region of positive $\omega_1$ on the high-latitude side and the region
of negative $\omega_1$ on the low-latitude side.  We are thus able to obtain an inward flow at
the latitudes where sunspots are typically seen, but unfortunately we are getting this at
the wrong time---shortly after the sunspot minimum (when there would be no sunspots in that region)
rather than at the sunspot maximum.  A look at Fig.~\ref{fig:bfly_1d12}(a) also makes it clear that this
inward flow occurs slightly after the sunspot minimum. We are now exploring the question
whether, by changing some parameters of the model, it is possible to get this inward flow
at the right latitudes at the right phase (around the sunspot maximum) of the solar cycle.
If our interpretation is correct, then the inward flow has nothing to do with the actual
physical presence of sunspots.  While the Lorentz force above the tachocline tends to
produce positive vorticity, the low-latitude edge of the toroidal magnetic field belt can
be a source of negative vorticity.  If the sunspot belt merely happens to be a region having
positive $\omega_1$ on the high-latitude side and negative $\omega_1$ on the low-latitude
side, then it will be a region of apparent inward flow.  We suggest this as a tentative
hypothesis which requires further study. We shall argue in Appendix~\ref{sec:appendix2} that the advection
term $s \nabla. ( \vb_1 \omega_0/s)$ not incorporated in the present study (the third
term on the LHS of Eq.~\ref{eq:main2}) may be quite important for modelling the inward
flow in active regions.

\begin{figure*}
\centering{
\includegraphics[width=0.84\textwidth]{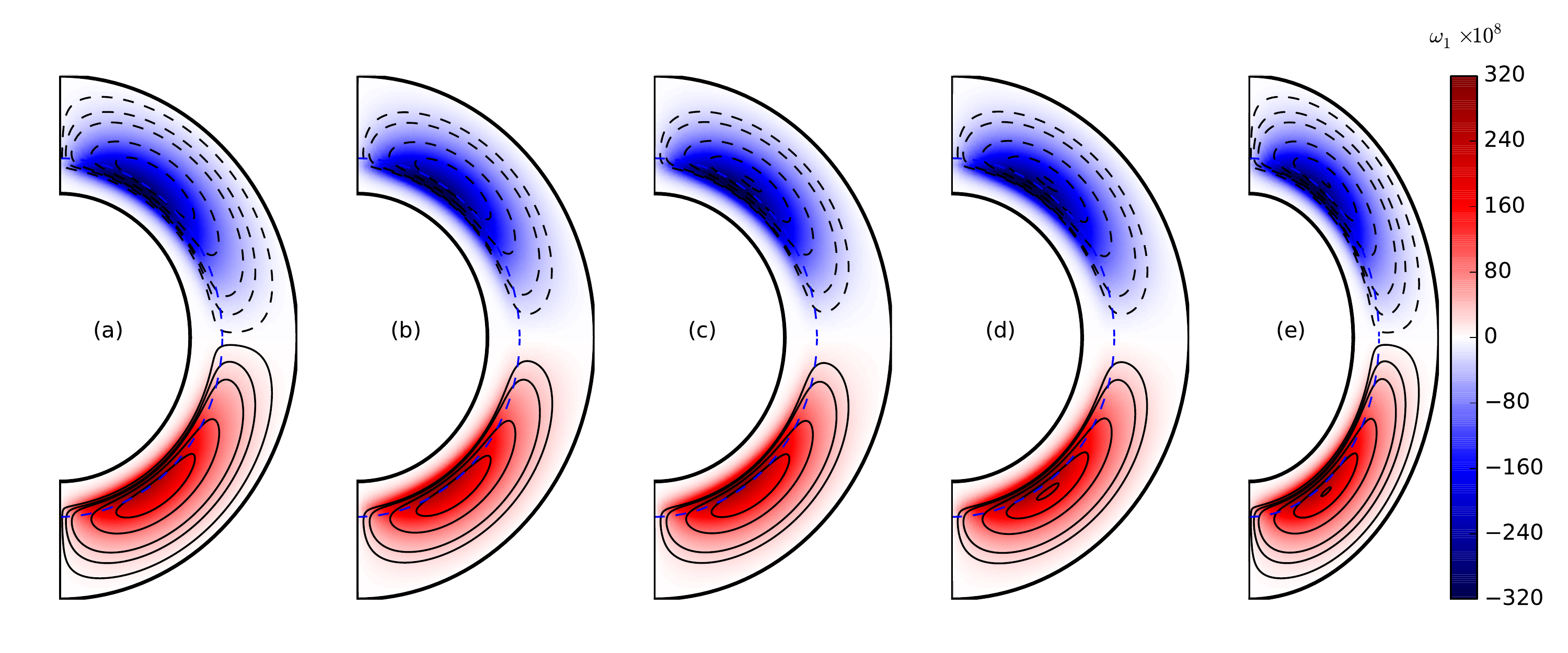}}
\caption{On changing the bottom boundary condition to $\omega_1 =0$ at $r = 0.55R_\odot$, 
the third row of Fig.~\ref{fig:snap_1d12} would get changed to this.}
\label{fig:snap_55}
\end{figure*}

\subsection{Results with low Prandtl number $(P_m < 1)$}\label{sec:lowpm}

In order to study how the nature of the perturbations in the \MC\ will change on
changing the turbulent viscosity $\nu$, we now present results obtained by taking
$\nu = \eta_t$ (shown by the blue solid line in Fig.~\ref{fig:eta}) rather than
$\nu = \eta_p$ which was the case for the results presented in Section~\ref{sec:pm_1}. This makes
the magnetic Prandtl number defined as $\nu/\eta_p$ (note that $\eta_p$ rather than
$\eta_t$ represents the unquenched turbulent diffusivity) much smaller than
unity within the body of the convection zone.  

We now want to present our results in the meridional 
plane of the Sun. Since there have been no changes in the dynamo equations ~\ref{eq:Aeq} and \ref{eq:Beq},
the toroidal field $B_\phi$ and the 
$\phi$-component of the curl of the Lorentz force $[\nabla \times {\bf F}_L]_{\phi}$ will
be the same as in the first two rows of Fig.~\ref{fig:snap_1d12} if we take the snapshots
at the same instants of time. To facilitate easy comparison with the perturbed vorticity
$\omega_1$ generated, we again plot $[\nabla \times {\bf F}_L]_{\phi}$ in the top row of
Figure~\ref{fig:snap_5d10}. This is the same as the middle row of Fig.~\ref{fig:snap_1d12}.
Then the bottom row of Figure~\ref{fig:snap_5d10} shows the perturbed vorticity $\omega_1$
along with the streamlines of the perturbed velocity.
This bottom row should be compared to the bottom row of Fig.~\ref{fig:snap_1d12}.
When $\nu$ is higher (the case of Fig.~\ref{fig:snap_1d12}), the perturbed vorticity
$\omega_1$ spreads more evenly within the convection zone.  On the other hand, when
$\nu$ is lower (the case of Figure~\ref{fig:snap_5d10}), the perturbed vorticity $\omega_1$
tends to be frozen where it is created by $[\nabla \times {\bf F}_L]_{\phi}$.  As a result,
$\omega_1$ follows $[\nabla \times {\bf F}_L]_{\phi}$ more closely in Figure~\ref{fig:snap_5d10}
rather than in Fig.~\ref{fig:snap_1d12}. We thus see a more complicated distribution
of $\omega_1$ in Figure~\ref{fig:snap_5d10} implying a more complicated
velocity field.

To make the perturbed meridional circulation near the surface 
comparable to what we find in the observational data, we need to choose $f = 5.3 \times 10^2$
in this case of low $P_m$. This is even more unphysical than what we needed for the case 
$P_m =1$ ($f = 53$) with all the other dynamo parameters same. Basically, on reducing
viscosity, vorticity tends to get piled up and we need to reduce the strength of the
Lorentz force further to match observations.
The perturbed velocities near the surface and 
near the bottom of the convection zone are shown in Fig.~\ref{fig:bfly_5d10}(a) and \ref{fig:bfly_5d10}(b) respectively. 
The requirement that the perturbed velocity at the surface compares with observational
data forces the perturbed velocity at the bottom of the convection zone 
as seen in Fig.~\ref{fig:bfly_5d10}(b) of order 1.5 m s$^{-1}$, which makes it almost comparable
to the unperturbed velocity (though they are in opposite directions). This shows that solving
the dynamo equations ~\ref{eq:Aeq} and \ref{eq:Beq} with the unperturbed meridional circulation
$\vb_0$ is more unjustified in this case than in the case of $P_m = 1$.

We do not know how large the perturbations to the \MC\ at the bottom of the convection
zone actually are. If they were comparable to the unperturbed \MC, then probably we would see some effects
on the dynamo.  Since this problem has not been studied before, we do not know what those
effects may be like.  As, on theoretical grounds also we expect $P_m$ to be of order unity,
we believe that the model with $P_m=1$ presented in Section~\ref{sec:pm_1} is probably a more realistic
depiction of what is happening inside the Sun rather than the model with low $P_m$.

\subsection{Dependence of the results on lower radial boundary condition}\label{sec:lowbc}

While discussing Fig.~\ref{fig:snap_1d12}, we have pointed out that the Lorentz force at the bottom of
the tachocline would have a tendency of producing a perturbed vorticity with the same
sign as the unperturbed vorticity. This will tend to drive a counter-clockwise flow (in the
northern hemisphere) below the tachocline.  Since on physical grounds we do not expect such a
flow to be driven in a region of stable temperature gradient, we have used the boundary
condition $\omega_1 = 0$ at $r = 0.70R_\odot$ in Sections~\ref{sec:pm_1} and \ref{sec:lowpm} to suppress this flow.
We found that our results are quite sensitive to this bottom boundary condition.  We now
discuss what we get on changing the boundary
condition to $\omega_1 = 0$ at $r = 0.55R_\odot$.

Again the solutions of the dynamo equations ~\ref{eq:Aeq} and \ref{eq:Beq} remain the same
as in Section~\ref{sec:pm_1} and \ref{sec:lowpm}. We present the solutions of the perturbed vorticity 
at the same time instants as the time instants
in Figure~\ref{fig:snap_1d12}. Only the third row of this figure will be 
replaced by what is shown in Figure~\ref{fig:snap_55}. When we had taken the boundary
condition $\omega_1 = 0$ at $r = 0.70R_\odot$ earlier, $\omega_1$ was restrained from
growing indefinitely. Now, on pushing the boundary well below the bottom of the convection
zone, we found that $\omega_1$ kept growing for many cycles for which we ran the code (see Fig.~\ref{fig:w1_time}).  This
is the reason why the values of $\omega_1$ given in the colour scale in Figure~\ref{fig:snap_55}
are rather large.  We would expect the growing perturbed vorticity to have the opposite
sign of the original unperturbed vorticity.  But this is not the case now.  The growing perturbed vorticity
has the same sign as the original unperturbed vorticity. This means that the Lorentz force
strengthens the original \MC\ rather than opposing it! Presumably this is because the effect of the Lorentz
force at the bottom of the tachocline (which tries to produce a perturbed vorticity of
the same sign as the original vorticity) overwhelms the effect of the Lorentz force at the top of the
tachocline which opposes the unperturbed \MC.  These results show that the bottom boundary
condition is quite crucial. If we allow the Lorentz force to create a perturbed vorticity
below the bottom of the tachocline, we may get totally unphysical results.

\section{Conclusion}\label{sec:conclusion}     

The \MC\ of the Sun plays a crucial role in the flux transport dynamo
model, the currently favoured theoretical model for the solar cycle.
Presumably an equatorward flow at the bottom of the convection zone is
responsible for the equatorward migration of the sunspot belts \citep{CSD95,HKC14}. 
The poleward flow at the surface builds up the polar field of the Sun by bringing
together the poloidal field generated by the Babcock--Leighton mechanism
from the decay of active regions \citep{HCM17}. 
We now have growing evidence that similar flux transport dynamos operate
in other solar-like stars as well \citep{Chou17}. Our lack of understanding
of the \MC\ in such stars is the major stumbling block for building theoretical
models of stellar dynamos \citep{KKC14, Chou17}.
The mathematical theory of the \MC\ involves a centrifugal term and a thermal wind
term.  The centrifugal term couples the theory of the \MC\ to the theory of
differential rotation and the thermal wind term couples the theory to the 
thermodynamics of the star---making it a very complex subject.

We now have evidence that the \MC\ of the Sun varies with the solar cycle, becoming
weaker around the sunspot maximum \citep{CD01,Beck02,Hathaway10b,Komm15}. 
We believe that this is caused by the back-reaction of the Lorentz force of the dynamo-generated
magnetic field. We show that the theory of the perturbations to the \MC\
caused by the Lorentz force can be decoupled from the theory of the 
unperturbed \MC\ itself.  Especially, this theory becomes decoupled from
any thermodynamic considerations if we assume that the magnetic fields do
not affect the thermodynamics of the Sun. This makes the theory of the
perturbations to the \MC\ much simpler than the theory of the \MC\ itself.
Even without having a theory of the unperturbed \MC, we are able to develop
a theory of how the \MC\ gets affected by the Lorentz force of the dynamo-generated
magnetic fields. 

The most convenient way of developing a theory of the \MC\ is by considering
the equation for the $\phi$-component of vorticity $\omega_{\phi}$. We solve
the equation for the perturbed vorticity along with the standard axisymmetric mean field equations of the
flux transport dynamo.  For the unperturbed \MC, $\omega_{\phi}$ is negative
in the northern hemisphere of the Sun.  If the Lorentz force can produce some
positive vorticity there, then we expect the \MC\ to become weaker during the
sunspot maxima. Our calculations show that positive vorticity is indeed produced
near the top of the tachocline.  Our theoretical results are able to explain
many aspects of observational data presented by \citet{Hathaway10b} and \citet{Komm15}
We get best results when the turbulent viscosity $\nu$
is taken to be comparable to the turbulent diffusivity $\eta_p$ of the magnetic
field, i.e.\ when the magnetic Prandtl number $P_m$ is close to unity.  Our
results are sensitive to the boundary condition at the bottom of the convection
zone. At the bottom of the tachocline, the Lorentz force tends to produce vorticity with
the same sign as the unperturbed vorticity and this has to be suppressed with a 
suitable boundary condition to incorporate the important physics that it is
very difficult to drive flows in the stable subadiabatic layers below the
tachocline. 

One intriguing observation is that of an inward flow towards active regions
during the sunspot maxima \citep{CS10, Komm15}. 
While we have not been able to construct a satisfactory model of this inward
flow, we suggest a possibility how this flow may be produced.  We would have
such a flow if the northern hemisphere has a positive perturbed vorticity
at latitudes above the mid-latitudes and a negative perturbed vorticity at lower
latitudes.  We propose a scenario how this can happen.  The Lorentz force at
the top of the tachocline tends to produce positive vorticity, whereas 
the Lorentz force at the edge
of the toroidal field belt at low latitudes tends to produce negative vorticity.
In our model, we found an inward flow towards typical sunspot latitudes
at a wrong phase of the solar cycle. It remains to be seen whether the phase
can be corrected by changing the parameters of the problem.

One puzzle we encounter is that we have to reduce the strength of the Lorentz
force artificially by about two orders of magnitude to make theory fit with
observations. We point out that a similar problem arises in the non-magnetic theory as well. 
While we do not have a proper explanation for this, we make
one comment.  While the overall strength of the Lorentz force varies with
the solar cycle, it does not change sign between cycles.  As a result, the 
Lorentz force would always tend to drive a very strong clockwise flow in the northern
hemisphere (anti-clockwise in the south), if it is not reduced sufficiently. 
Now, in the theory of the unperturbed \MC, we know that we need some force
trying to drive such a flow---to make the surfaces of constant angular velocity
depart from cylinders and to balance the centrifugal term.  Normally, the thermal
wind caused by anisotropic heat transport in the Sun is believed 
to do this \citep{Kitchatinov95,Kitchatinov11b}. Is it possible that the expected large Lorentz force
also plays an important in the force balance?  To address this question, one
has to carry on a much more complicated analysis than what we have attempted
in this paper, without separating the equations of unperturbed and perturbed \MC.
While we do not attempt such an analysis, our analysis brings out many basic
physics issues rather clearly and raises intriguing questions like this.

The equation of the perturbed \MC\ gives us the perturbed vorticity at different
times.  In order to get the perturbed velocity from the perturbed vorticity,
we need to solve a Poisson-type equation. If we want to do this at every time
step, then running the code becomes computationally very expensive.  On the hand,
if we are satisfied only with the perturbed vorticity at all steps and do not
require the perturbed velocity, then we need much less computer time.  This is what
we have done in this paper in which our aim was to carry out a parameter space
study to understand the basic physics.  However, we have to pay a price for not
calculating the perturbed velocity at all time steps.  We keep solving the
dynamo equations by using the unperturbed velocity of the \MC. This is certainly a
questionable approximation. If we want to make the perturbed velocity at the
surface comparable to what is found in observational data ($\approx$ 5 m s$^{-1}$),
then the perturbed velocity at the bottom of the convection zone turns out to
be $\approx$ 0.75 m s$^{-1}$. This is certainly smaller than the unperturbed
velocity there ($\approx$ 2 m s$^{-1}$), but is not completely negligible.
While we feel reassured that the perturbed velocity at the bottom of the
convection zone is smaller than the unperturbed velocity and probably will
not necessitate a major revision of the flux transport dynamo, the effect of this
perturbed velocity on the dynamo needs to be investigated.
We are now in the process of including this perturbed velocity in the
dynamo equations.  This will need finding the perturbed velocity from the
perturbed vorticity at different time intervals---pushing up the computational
requirements significantly. Once we do that, we shall be able to include another
advection term involving ${\bf v}_1$ in the computations which had been left
out in this paper because of computational difficulties.  This term may be
important for modelling the inward flow towards active regions.

Not including the back-reactions on the \MC\ in the dynamo equations is not
a limitation of this paper alone, but a limitation of nearly all hitherto
published papers on the flux transport dynamo based on 2D kinematic mean
field formulation.  To the best of our knowledge, only \citet{KarakChou12}
studied this problem by modelling the back-reaction on the \MC\
due to the Lorentz force through a simple parameterization (though see
\citet{PCB12} who have included back-reaction in a simple low-order dynamo model).  
Although such a back-reaction on the \MC\ can have dramatic effects on the dynamo if the
magnetic diffusivity is low, \citet{KarakChou12} found that the effect
is not much on a high-diffusivity dynamo (like what we present in this paper).
However, it is now important to go beyond such simple parameterization and
include the properly computed perturbed velocity in the dynamo equations and
study its effects.  We hope that we shall be able to present our results in
a forthcoming paper in the near future.

\section*{Acknowledgements}
We thank an anonymous referee for valuable comments. GH would like to acknowledge CSIR, India for financial support. ARC's research is supported
by J. C. Bose fellowship granted by DST, India. 

\bibliographystyle{mnras}
\bibliography{myref}
\appendix
\section{Calculation of the Viscosity Term}\label{sec:appendix}

The calculation of the viscous term becomes immensely complicated if
$\nabla. {\vb}$ is not set equal to zero.  Although we are dealing with
a situation in which we rather have $\nabla. (\rho {\vb}) = 0$, the
meridional circulation velocities are everywhere highly subsonic and
the role of compressibility in viscous dissipation is expected to be
negligible.  So we shall assume $\nabla. {\vb} = 0$ to make the calculations
manageable.  For a velocity field 
$$\vb = v_r (r, \theta,t) {\bf e}_r + v_{\theta} (r, \theta,t) {\bf e}_{\theta}$$
the non-zero components of the viscous stress tensor are
\begin{eqnarray}
\sigma_{rr} = 2 \nu \frac{\pa v_r}{\pa r}, \sigma_{\theta \theta}
= 2 \nu \left( \frac{1}{r} \frac{\pa v_{\theta}}{\pa \theta} 
+ \frac{v_r}{r} \right), ~~~~~~~~~~~~~~~~~~~~~~\nonumber\\
\sigma_{\phi \phi}= 2 \nu \left( \frac{v_r}{r}
+\frac{v_{\theta} \cot \theta}{r} \right), 
\sigma_{r \theta} = \nu \left(\frac{1}{r} \frac{\pa v_r}{\pa \theta}
+ \frac{\pa v_{\theta}}{\pa r} - \frac{v_{\theta}}{r} \right)
\label{eq:A1}
\end{eqnarray}
The $r$ and $\theta$ components of the viscosity term in Navier--Stokes
equation are given by
\begin{equation}
F_{\nu,r} = \frac{1}{r^2} \frac{\pa}{\pa r} (r^2 \sigma_{rr}) +
\frac{1}{r \sin \theta} \frac{\pa}{\pa \theta}(\sin \theta \sigma_{r \theta})
-\frac{\sigma_{\theta \theta} + \sigma_{\phi \phi}}{r}, 
\label{eq:A2}
\end{equation}
\begin{equation}
F_{\nu,\theta} = \frac{1}{r^2} \frac{\pa}{\pa r} (r^2 \sigma_{r \theta}) +
\frac{1}{r \sin \theta} \frac{\pa}{\pa \theta}(\sin \theta \sigma_{\theta \theta})
+ \frac{\sigma_{r \theta}}{r} - \frac{\sigma_{\phi \phi} \cot \theta}{r}, 
\label{eq:A3}
\end{equation}
We now substitute from Eq.~\ref{eq:A1} in Equations~\ref{eq:A2} and \ref{eq:A3}. On making use of 
$\nabla. {\vb} = 0$, we get after a few steps of algebra
\begin{equation}
F_{\nu,r} = \nu \left[ \nabla^2 v_r - \frac{2}{r^2} 
\frac{\pa v_{\theta}}{\pa \theta}
- \frac{2 v_r}{r^2} - \frac {2 \cot \theta v_{\theta}}{r^2} \right]
+ 2 \frac{d \nu}{d r} \frac{\pa v_r}{\pa r},
\label{eq:A4}
\end{equation}
\begin{equation}
F_{\nu,\theta} = \nu \left[ \nabla^2 v_{\theta} + \frac{2}{r^2} 
\frac{\pa v_r}{\pa \theta}
- \frac { v_{\theta}}{r^2 \sin^2 \theta} \right]
+ \frac{d \nu}{d r} \left[\frac{1}{r} \frac{\pa v_r}{\pa \theta}
+ \frac{\pa v_{\theta}}{\pa r} - \frac{v_{\theta}}{r} \right].
\label{eq:A5}
\end{equation}

To solve Equation~\ref{eq:main2}, we need $[\nabla \times {\bf F}_{\nu}]_{\phi}$, which
is given by
\begin{equation}
[\nabla \times {\bf F}_{\nu}]_{\phi} = \frac{1}{r}
\left[ \frac{\pa}{\pa r}(r F_{\nu, \theta}) - \frac{\pa F_{\nu, r}}
{\pa \theta} \right]. 
\label{eq:A6}
\end{equation}
\begin{figure}
\centering{
\includegraphics[width = 0.5\textwidth,trim={0 0.75cm 0 0}]{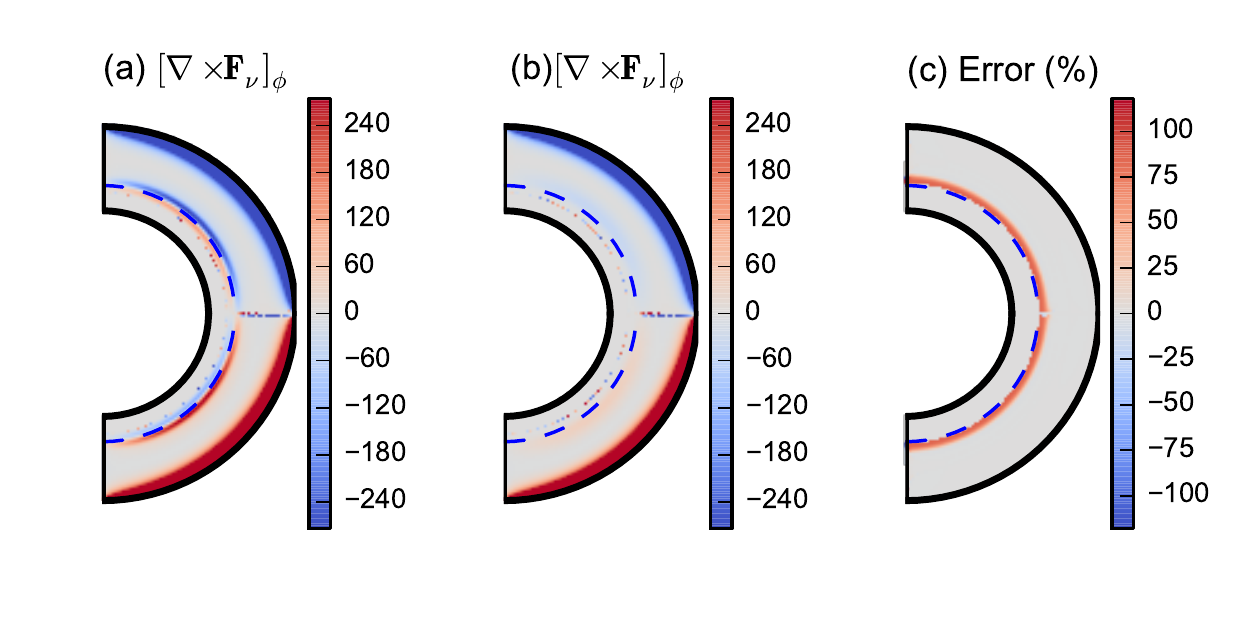}}
\caption{$[\nabla \times {\bf F}_{\nu}]_{\phi}$ calculated using full expression (Eq.~\ref{eq:A6}) and using approximated expression (Eq.~\ref{eq:viscous1}) are shown
in (a) and (b) respectively. The difference between the values of case (a) and (b) are shown in percentage in (c).}   
\label{fig:v_force}
\end{figure}
On substituting Equations~\ref{eq:A4} and \ref{eq:A5} into Eq.~\ref{eq:A6}, we get a rather complicated
analytical expression for the viscosity term.  It is extremely challenging
to incorporate this complicated expression in the time evolution code to
solve Eq.~\ref{eq:main2}---especially if we want to handle the diffusion terms through
Crank-Nicholson scheme which is partly implicit, to avoid very small time steps. 
Since there are many uncertainties in this problem---including
the uncertainty in the value of turbulent viscosity $\nu$---we felt that
it is not worthwhile to make an enormous effort to incorporate a completely
accurate expression of the viscous term in our code to solve Eq.~\ref{eq:main2}. In order
to capture the effect of viscosity in the problem, we have used the
simpler approximate expression given in Eq.~\ref{eq:viscous1}. The question is whether
we make a very large error in doing this simplification.  To address this
question, we take the unperturbed meridional circulation shown in Fig.~\ref{fig:stream_vort}(a)
and calculate $[\nabla \times {\bf F}_{\nu}]_{\phi}$ for this velocity
field both by the approximate expression in Eq.~\ref{eq:viscous1} and the full expression
in Eq.~\ref{eq:A6}.  To calculate the value of $[\nabla \times {\bf F}_{\nu}]_{\phi}$ at
different grid points by the full expression in Eq.~\ref{eq:A6}, we first calculate
$F_{\nu,r}$ and $F_{\nu, \theta}$ at different grid points by using
Equations~\ref{eq:A4} and \ref{eq:A5}. Then we use these numerical values of 
$F_{\nu,r}$ and $F_{\nu, \theta}$ to calculate 
$[\nabla \times {\bf F}_{\nu}]_{\phi}$ by using a simple difference
scheme to handle Eq.~\ref{eq:A6}. The values obtained for
$[\nabla \times {\bf F}_{\nu}]_{\phi}$ by using the full expression (Eq.~\ref{eq:A6}) and 
the approximate expression (Eq.~\ref{eq:viscous1}) are shown in 
Figure~\ref{fig:v_force}(a) and Figure~\ref{fig:v_force}(b) respectively. We
see that both the expressions give fairly close values within
the main body of the convection zone.  Only at the bottom of the convection
zone where radial derivatives of $\nu$ included in the full expression
are important, the two expressions give somewhat different values---though
the values are of the same order of magnitude (see Fig~\ref{fig:v_force}(c)). In Figure~\ref{fig:v_force}(c), we have 
shown percentage differences of $[\nabla \times {\bf F}_{\nu}]_{\phi}$ values calculated in 
two cases (Figures~\ref{fig:v_force}(a) and (b)) in our integration region 
above $r > 0.7R_\odot$. Note that things change
much more at the bottom of the convection zone on changing the bottom
boundary condition slightly. 
We thus conclude that,
by using the approximate expression in Eq.~\ref{eq:viscous1} for
$[\nabla \times {\bf F}_{\nu}]_{\phi}$, we do not make a large
error and get everything within the correct order of magnitude.

\section{Importance of various  advective terms}\label{sec:appendix2}
\begin{figure}
\centering{
\includegraphics[width = 0.5\textwidth,trim={0 0.75cm 0 0}]{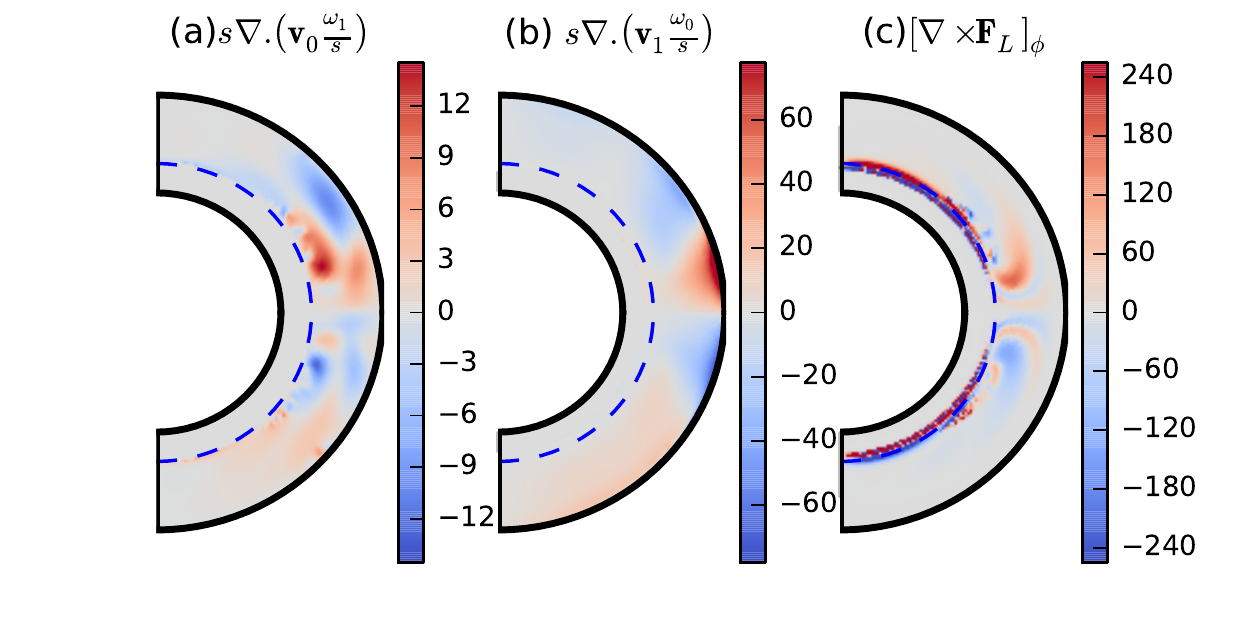}}
\caption{The advective terms in LHS of Eq.~\ref{eq:main2} are shown in (a) and (b), where $s=r\sin\theta$.
The $\phi$ component of the curl of the Lorentz force is shown in (c). All terms are calculated
at a particular time step near a solar maximum.}   
\label{fig:advect}
\end{figure}

In the calculations presented in this paper, 
we have neglected the term $s\nabla.\left({\bf v}_1\frac{\omega_0}{s}\right)$ while solving
Eq.~\ref{eq:main2}, because its evaluation requires the perturbed velocity ${\bf v}_1$ in every time 
step which is computationally
very expensive to obtain. Now we calculate the two advective terms in Eq.~\ref{eq:main2} separately using simple finite 
difference scheme at a particular time step near a solar maximum to see their relative importance
with respect to the Lorentz force 
term $[\nabla \times {\bf F}_L]_{\phi}$. The values of the two advection terms involving 
unperturbed velocity $s\nabla.\left({\bf v}_0\frac{\omega_1}{s}\right)$ 
and involving perturbed velocity $s\nabla.\left({\bf v}_1\frac{\omega_0}{s}\right)$  in the $r - \theta$ plane
are shown in Fig.~\ref{fig:advect}(a) and 
\ref{fig:advect}(b) respectively. The $\phi$ component of the curl of the Lorentz force $[\nabla \times {\bf F}_L]_{\phi}$ 
is shown in Fig.~\ref{fig:advect}(c). The colour table for curl of the Lorentz force is clipped to 240 to show its value in 
the convection zone (where it has much smaller values compared to the tachocline) but its maximum value is about 3000. 
Thus, in comparison to the curl of the Lorentz force, the other terms are small but not negligible. 
The advective term $s\nabla.\left({\bf v}_0\frac{\omega_1}{s}\right)$ which we already included
is important within the convection zone, but the other term with perturbed 
velocity $s\nabla.\left({\bf v}_1\frac{\omega_0}{s}\right)$ which we plan to include in our future paper
is important near the surface at low latitudes. 
The dynamics of vorticity presented in the Section~\ref{sec:pm_1} clearly reflect the 
fact that the time evolution of vorticity is mostly determined by 
the Lorentz force term $[\nabla \times {\bf F}_L]_{\phi}$ (see Fig.~\ref{fig:snap_1d12}((k)-(o))) even
when the advective term $s\nabla.\left({\bf v}_0\frac{\omega_1}{s}\right)$ is included. However, since the 
$s\nabla.\left({\bf v}_1\frac{\omega_0}{s}\right)$ term becomes appreciable in the regions where we have
inward flow towards active regions during the sunspot maximum, this term may have some importance
in modelling this inward flow. We plan to investigate this further in future.

\label{lastpage}
\end{document}